\documentclass{aa}  

\usepackage{graphicx}
\usepackage{orcidlink}
\usepackage{multirow}
\usepackage{amsmath}
\usepackage{amssymb}
\usepackage{float}
\usepackage{enumitem}
\usepackage{capt-of}

\usepackage{txfonts}

\newcommand{\kms}{$\textrm{km~s$^{-1}$}$}
\newcommand{\oi}{\,{\sc i}}
\newcommand{\ii}{\,{\sc ii}}
\newcommand{\iii}{\,{\sc iii}}
\newcommand{\HI}{H\,\textsc{i}}
\newcommand{\Ha}{H$\alpha$}
\newcommand{\Hb}{H$\beta$}
\newcommand{\Hg}{H$\gamma$}

\newcommand{\nb}{\textsc{NBursts}}

\newcommand{\Msun}{M\ensuremath{_\odot}\,}
\newcommand{\Lsun}{L\ensuremath{_\odot}\,}

\definecolor{MyOrange}{RGB}{255,165,0} 
\definecolor{MyPurple}{RGB}{128,0,128} 
\definecolor{MyMagenta}{RGB}{255,0,255} 
\graphicspath{{./}{}}

\begin{document} 

   \title{CoRGI in MaNGA -- II. NGC~5717: cold-gas accretion and disk growth in a BGG}
   
   \titlerunning{NGC~5717: BGG with stellar counter-rotation}

   \author{
      Damir~Gasymov\inst{1,2}\orcidlink{0000-0002-1750-2096}\corrauth{damir.gasymov@stud.uni-heidelberg.de}
      \and Ivan~Yu.~Katkov\inst{3,4,2}\orcidlink{0000-0002-6425-6879}
      \and Eva~K.~Grebel\inst{1}\orcidlink{0000-0002-1891-3794}
      \and Anna Pasquali\inst{1}\orcidlink{0000-0001-5171-5629}
      \and Evgenii~V.~Rubtsov\inst{2}\orcidlink{0000-0001-8427-0240}
      \and Katarina~Kraljic\inst{5}\orcidlink{0000-0001-6180-0245}
      \and Igor~V.~Chilingarian\inst{6}\orcidlink{0000-0002-7924-3253}
      \and Kirill~A.~Grishin\inst{7,2}\orcidlink{0000-0003-3255-7340}
      \and Anna~S.~Saburova\inst{2}\orcidlink{0000-0002-4342-9312}
    }
    
    \institute{
      Astronomisches Rechen-Institut, Zentrum für Astronomie der Universität Heidelberg, 
      Mönchhofstr. 12--14, 69120 Heidelberg, Germany
      \and Sternberg Astronomical Institute, Lomonosov Moscow State University, 
      Universitetskij pr. 13, 119234 Moscow, Russia
      \and New York University Abu Dhabi, PO Box 129188, Abu Dhabi, UAE
      \and Center for Astrophysics and Space Science (CASS), 
      New York University Abu Dhabi, PO Box 129188, Abu Dhabi, UAE
      \and Observatoire Astronomique de Strasbourg, UMR 7550, CNRS, Université de Strasbourg, F-67000 Strasbourg, France
      \and Center for Astrophysics --- Harvard and Smithsonian, 60 Garden St. MS09, Cambridge, MA 02138, USA
      \and Universite Paris Cite, CNRS, Astroparticule et Cosmologie, F-75013 Paris, France
    }

   \date{Received XXX, YYY; accepted XXX, YYY}
 
  \abstract
  {
    Large-scale stellar counter-rotation, in which two coplanar stellar components rotate in opposite directions, is usually interpreted as a signature of external gas accretion.
    Such systems are most often found in relatively sparse environments, where cold gas can be acquired and remain dynamically undisturbed long enough to settle into a disk.
    It is less clear whether this channel can operate in massive central galaxies, whose group or cluster environments can heat, strip, or perturb cold gas.
    Here we present NGC~5717, one of the most massive galaxies known to host a counter-rotating stellar disk and the first identified case in a brightest group galaxy.
    Using integral-field spectroscopy from SDSS-IV Mapping Nearby Galaxies at Apache Point Observatory (MaNGA), together with CFHT imaging, ancillary H\,\textsc{i} data, and SDSS/DESI spectroscopy of the surrounding group, we show that the galaxy contains a dynamically cold counter-rotating disk embedded in an old dynamically hot elliptical component.
    The disk is young ($\sim 1.3$~Gyr), metal-poor ([Fe/H] $\sim -0.7$ dex), extended ($h\sim2.4$~kpc), and massive ($\log M_\star/M_\odot \sim 9.9$), the associated ionized gas co-rotates with it, while the available H\,\textsc{i} data indicate no significant remaining reservoir of neutral atomic gas ($\log M_{\mathrm{HI}}/M_\odot \lesssim 9$).
    These properties support an \textit{in situ} origin of the secondary component from externally accreted gas.
    The chemical evolution model favors gas accretion from a cosmic filament, which is consistent with the surrounding cosmic-web geometry, and indicates highly efficient burst-like star formation with substantial metal-enriched outflows.
    NGC~5717 may therefore represent a rare example of disk growth in a massive central galaxy, enabled by external cold-gas accretion during an earlier phase of the group assembly history.
    }

   \keywords{galaxies: disk -- galaxies: kinematics and dynamics -- galaxies: stellar content -- galaxies: formation}

   \maketitle
   \nolinenumbers

\section{Introduction}

Kinematically decoupled gaseous and stellar components provide important tracers of external accretion and galaxy assembly. 
In particular, misaligned, polar, and counter-rotating (CR) gas disks are commonly interpreted as the result of externally acquired cold gas, which may subsequently settle and form secondary stellar components \citep{Davis2011MNRAS.417..882D,Corsini2014ASPC..486...51C,Katkov2014MNRAS.438.2798K}.
Large samples of such systems make it possible to constrain the origin of the accreted material and the physical conditions under which it can survive and form stars \citep{Moiseev2011MNRAS.418..244M,Young2020MNRAS.495.1433Y,Gasymov2025ApJS..281...19G}.

NGC~5717 stands out in our ``Stellar Counterrotating Galaxies Identified in the MaNGA Survey'' sample (CoRGI in MaNGA; Sloan Digital Sky Survey IV Mapping Nearby Galaxies at Apache Point Observatory, SDSS-IV MaNGA)  of galaxies with counter-rotating stellar disks \citep{Gasymov2025ApJS..281...19G} as an exceptional system.
Unlike most galaxies with stellar counter-rotation, it appears to be a massive elliptical galaxy ($\log M_\star/\Msun = 11.20$; Table~\ref{tab1}) hosting an embedded large-scale counter-rotating stellar disk within an old, weakly rotating spheroidal body.
Although this morphological classification differs from those adopted in other catalogs (S0--Sa; \citealt{Vazquez-Mata2025MNRAS.544.1056V}), we demonstrate the spheroidal nature of the dominant component in Appendix~\ref{app:morph_mismatch}.
A cross-match with the group catalog of \citet{Yang2007ApJ...671..153Y} further shows that NGC~5717 is the brightest galaxy in its host group.
The formation of a young, dynamically cold counter-rotating disk in such a massive central galaxy is therefore challenging, as it requires cold gas to be acquired, survive the group environment, settle into an ordered configuration, and form stars.

The role of environment is important to galaxy evolution.
Galaxy properties are known to correlate with local density and halo mass, from the classical morphology--density relation to the distinct assembly histories of central and satellite galaxies \citep{Dressler1980ApJ...236..351D,Peng2010ApJ...721..193P,Somerville2015ARA&A..53...51S}.
These correlations extend to the most massive central galaxies: the properties of brightest cluster and group galaxies (BCGs and BGGs) are connected to the global properties and dynamical state of their host groups and clusters, including the mass, shape, and orientation of the surrounding halo.

Cold-gas accretion is a key ingredient of galaxy evolution, but its observability and efficiency depend strongly on environment \citep{Sancisi2008A&ARv..15..189S,Tacconi2020ARA&A..58..157T}.
In relatively sparse environments, gas accreted from outside the galaxy can remain for long periods and settle into extended disks \citep{Oosterloo2007A&A...465..787O,Katkov2014MNRAS.438.2798K}.
In groups and clusters, by contrast, the survival of cold gas in galaxies may be limited by strangulation, ram-pressure stripping, heating by the hot intra-group or intra-cluster medium, AGN feedback, and repeated tidal encounters \citep{Gunn1972ApJ...176....1G,Moore1996Natur.379..613M,Boselli2006PASP..118..517B,Boselli2022A&ARv..30....3B}.

Here, we investigate the properties and origin of the counter-rotating disk in NGC~5717, which is fully covered by the MaNGA field of view.
We determine the stellar-population properties of the main and secondary components, analyze the ionized and atomic H\,\textsc{i} gas, characterize the host group environment, and construct a chemical evolution model to constrain the gas infall and star-formation history of the counter-rotating disk.

This paper is organized as follows.
In Section~\ref{sec:data}, we describe the data and catalogs used in this work.
In Section~\ref{sec:methods}, we describe the data-analysis procedure.
In Section~\ref{sec:discussion}, we discuss the properties of the star-formation process in the galaxy, the origin of the gas, and the rarity of such systems.
Throughout the paper, we adopt the WMAP9 cosmology with $H_0~=~69.3$~km s$^{-1}$ Mpc$^{-1}$ and $\Omega_0 = 0.2865$ \citep{Hinshaw2013ApJS..208...19H}.
Unless stated otherwise, all uncertainties are quoted at the $1\sigma$ level and all logarithms are base 10.
Some catalogs adopt $h = 0.7$ \citep{Yang2007ApJ...671..153Y, Stark2021MNRAS.503.1345S}, or $h = 0.75$ \citep{Zhang2024SCPMA..6719511Z}, but we assume that the difference is negligible for the purposes of this work.
Throughout this paper, we refer to the embedded counter-rotating disk as the secondary (young, cold, central, CR) component or structure and to the surrounding elliptical body as the main (old, hot, spheroidal, host) component.

\section{Data description}
\label{sec:data}

\begin{figure*}
\centering
\includegraphics[width=0.98\textwidth]{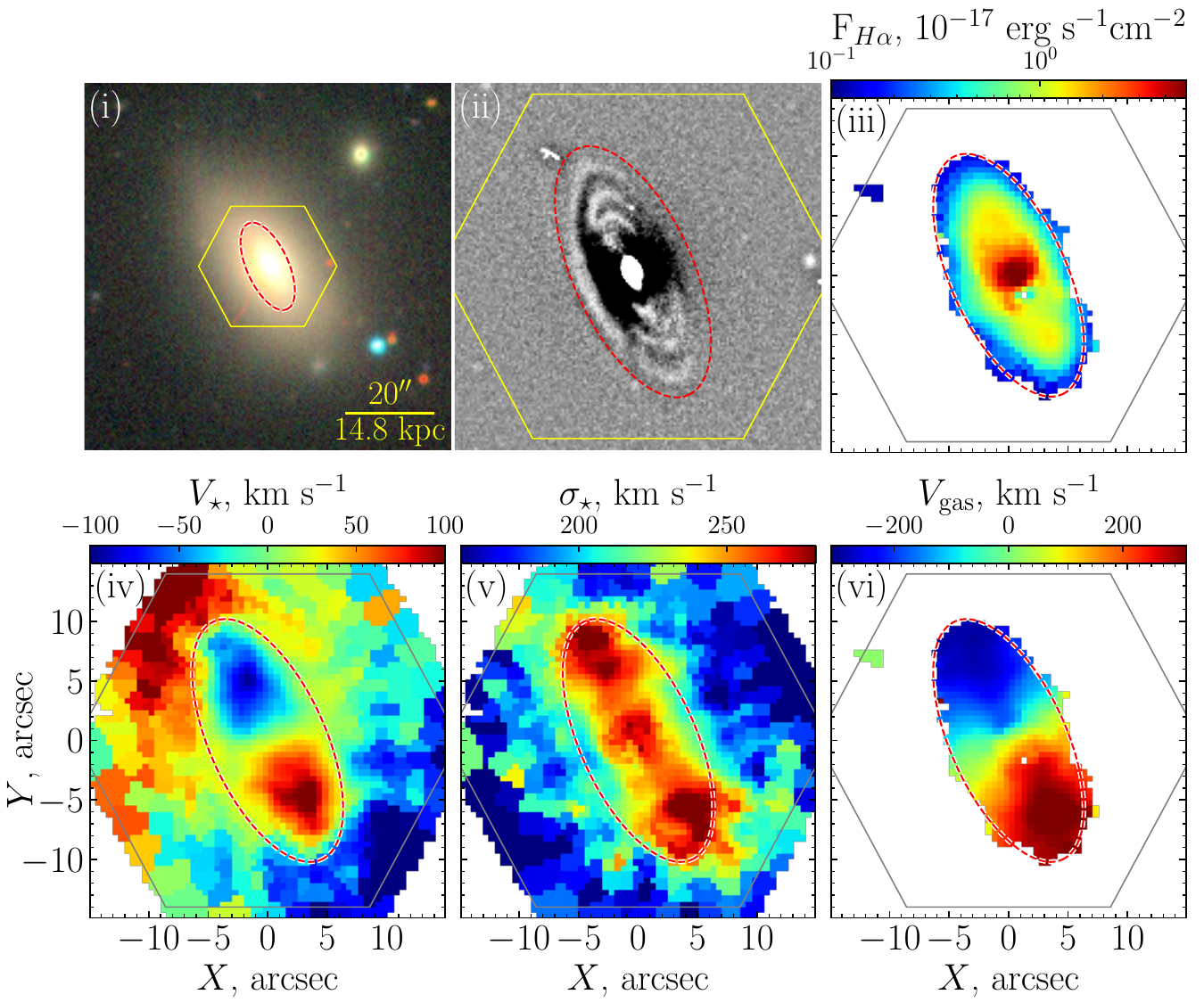}
\caption{Photometric and kinematic properties of NGC~5717. 
From left to right and top to bottom: (i) a composite RGB image from the Legacy Imaging Surveys \citep{Dey2019_legacysurveys}; the yellow and gray hexagons indicate the MaNGA fields of view, and the red ellipses trace the visual boundary of the embedded CR disk, whose semi-major axis is 11 arcsec, or 8 kpc;
(ii) an unsharp-masked image based on the CFHT $r$-band data, highlighting the three-ring structure of the disk; 
(iii--vi) maps derived from the one-component \nb\ analysis (Sect.~\ref{subsec:kin_fit}): \Ha\ flux $F_{\mathrm{H}\alpha}$, stellar line-of-sight velocity $V_\star$, stellar velocity dispersion $\sigma_\star$, and ionized-gas velocity $V_{\mathrm{gas}}$.
For the $F_{\mathrm{H}\alpha}$ and $V_{\mathrm{gas}}$ maps, only spaxels with a signal-to-noise ratio greater than 5 in the H$\alpha$ emission line are shown.
\label{fig:kin_maps}}
\end{figure*}

The spectral data for the galaxy were taken from the MaNGA survey \citep{Abdurro'uf2022ApJS..259...35A} based on integral-field-unit (IFU) spectroscopy \citep{Drory2015AJ....149...77D}, conducted with the SDSS 2.5-m telescope \citep{Gunn2006AJ....131.2332G} at Apache Point Observatory, of about 10,000 galaxies in the nearby Universe ($0.01~\lesssim~z~\lesssim~0.15$) covering a wide range of stellar mass and colors \citep{Wake2017AJ....154...86W}.
In the survey, each galaxy was observed using 2~arcsec fibers packed into bundles with diameters ranging from 12~arcsec (19 fibers) to 32~arcsec (127 fibers), covering out to $(1.5-2.5) R_{\rm eff}$, where $R_{\rm eff}$ is the effective radius of the target.
For NGC~5717, the largest bundle was used, and it covers 1.6 $R_{\rm eff}$ ($R_{\rm eff} = 10.2''$ or $7.7$~kpc).
Fiber bundles feed the Baryon Oscillation Spectroscopic Survey (BOSS) spectrographs \citep{Smee2013AJ....146...32S}, which provide a wavelength coverage of $\lambda\lambda$3600--10300~\AA\ with a spectral resolution $R\sim2000$, corresponding to an instrumental dispersion $\sigma_\mathrm{inst.}\approx75$~\kms\ at 5100~\AA\ \citep{Law2016AJ....152...83L}.
To fill in the gaps between the round fibers in a bundle, each galaxy was observed using 3 spatially dithered exposures \citep{Law2015AJ....150...19L, Yan2016AJ....152..197Y}.

Besides spectral data, we used values from MaNGA Value-Added Catalogs (VACs) and external galaxy catalogs:
\begin{itemize}
    \item The MaNGA-HI DR4 \citep{Masters2019MNRAS.488.3396M, Stark2021MNRAS.503.1345S} provides atomic hydrogen measurements based on 21~cm observations from the Green Bank Telescope (GBT), complemented by the ALFALFA survey \citep{Haynes2018ApJ...861...49H}. For NGC~5717, a detection is reported based on a GBT radio spectrum with an FWHM $\approx 9'$ beam;
    \item The MaNGA Visual Morphology catalog (MVM-VAC) \citep{Vazquez-Mata2025MNRAS.544.1056V} provides visual morphological classification for MaNGA galaxies. Reliable classifications are available for 9361 galaxies. NGC~5717 is classified as an Sa-type galaxy;
    \item The NASA-Sloan Atlas (NSA) \citep{Blanton2011AJ....142...31B, Wake2017AJ....154...86W} contains photometric and structural parameters for nearby galaxies based on SDSS imaging;
    \item The Reference Catalog of Spectral Energy Distributions (RCSED) \citep{Chilingarian2017ApJS..228...14C} offers homogenized UV-NIR photometry and SDSS spectroscopic measurements and analysis for $\sim$800,000 galaxies;
    \item The Galaxy Groups catalog \citep{Yang2007ApJ...671..153Y}, which is based on SDSS DR4 data. The objects were selected in the redshift range $0.01 \lesssim z \lesssim 0.20$ with a redshift completeness $C > 0.7$;
    \item The FAST All Sky HI survey (FASHI) \citep{Zhang2024SCPMA..6719511Z} provides blind 21~cm \HI\ measurements from the Five-hundred-meter Aperture Spherical radio Telescope (FAST).
\end{itemize}

Deep optical imaging of NGC~5717 was retrieved from the Canadian Astronomy Data Centre (CADC) archive, which provides access to data products from multiple facilities including the Canada-France-Hawaii Telescope (CFHT). 
The observations were obtained with the MegaPrime/MegaCam wide-field imager on CFHT \citep{Boulade2003SPIE.4841...72B}. 
The CADC archive contains nine MegaCam exposures of NGC~5717: six in the $u$-band (CFHT/u.MP9302) and three in the $r$-band (CFHT/r.MP9602); nominal $3\sigma$ surface-brightness limit $\mu_r \simeq 27.9$ mag arcsec$^{-2}$.

The 1D spectral data for galaxies in and around the NGC~5717 group were retrieved from the Sloan Digital Sky Survey Data Release 17 (SDSS DR17) \citep{York2000AJ....120.1579Y, Abdurro'uf2022ApJS..259...35A} and from the Dark Energy Spectroscopic Instrument (DESI) Data Release 1 survey \citep{DESICollaboration2026AJ....171..285D}. 
SDSS optical spectroscopy was obtained with plug-plate observations covering a $\sim$3$^\circ$ diameter field; depending on the SDSS program, targets are assigned to either the original SDSS spectrographs (640 fibers of 3\arcsec\ diameter) or to the BOSS spectrographs (1000 fibers of 2\arcsec\ diameter), providing wavelength coverage of 3800--9200~\AA\ (SDSS) and 3650--10400~\AA\ (BOSS) at $R\sim1500$--2500 \citep{York2000AJ....120.1579Y, Abdurro'uf2022ApJS..259...35A}. DESI is a fiber-fed spectroscopic survey conducted on the Mayall 4-m telescope using 1.5\arcsec\ diameter fibers, providing continuous wavelength coverage from 3600-9800\AA\ at a spectral resolution of $R\sim2000$--5500 \citep{DESICollaboration2016arXiv161100036D, DESICollaboration2026AJ....171..285D}.

\section{Methods and analysis}
\label{sec:methods}

\begin{table}
\caption{Properties of NGC~5717 and its host group.}
\label{tab1}
\centering
\begin{tabular}{ll}
\hline\hline
Parameter & NGC~5717 \\
\hline
RA (J2000.0) & 14:38:37.7 \\
Dec. (J2000.0) & $+46$:39:47.23 \\
$z_{\rm gal}$ & 0.03689 \\
$\log M_\star/\Msun$ & 11.20 \\
Distance, Mpc & 158.3 \\
Scale, kpc arcsec$^{-1}$ & 0.740 \\
Morphological type & S0-a \\
Kinematic features & 2$\sigma$, CR-GS, Srot \\
\hline
Parameter & NGC~5717 group \\
\hline
RA (J2000.0) & 14:38:28.0 \\
Dec. (J2000.0) & $+46$:41:32.82 \\
$z_{\rm cl}$ & 0.03638 \\
$N_{\rm mem}$ & 27 \\
$\log L_r/\Lsun$ & 11.08 \\
$\log M_\star/\Msun$ & 11.55 \\
$\log M_{\rm halo}/\Msun$ & 13.72 \\
$R_{200}(M_{\rm halo})$, Mpc & 1.31 \\
$\sigma_{\rm cl}$, \kms & 209 \\
$\sigma_{\rm cl}(M_{\rm halo})$, \kms & 323 \\
\hline
\end{tabular}
\tablefoot{
The galaxy coordinates, redshift, stellar mass, distance, physical scale, morphological type, and the observed kinematic features are listed in the upper part of the table (descriptions are given by \citet{Gasymov2025ApJS..281...19G}), while the bottom half summarizes the properties of the galaxy group from the catalog of \citet{Yang2007ApJ...671..153Y}. 
For the group, $\sigma_{\rm cl}$ is the velocity dispersion estimated from the identified group members, while $\sigma_{\rm cl}(M_{\rm halo})$ and $R_{200}(M_{\rm halo})$ are the velocity dispersion and $R_{200}$ expected from the halo mass \citep{Yang2007ApJ...671..153Y, Pasquali2019MNRAS.484.1702P}.}
\end{table}

\subsection{Kinematic model of the galaxy}
\label{subsec:kin_fit}

To estimate the stellar masses of the components in NGC~5717, we follow the standard workflow previously applied to galaxies with counter-rotating (CR) stellar disks, including IC~719 \citep{Katkov2013ApJ...769..105K}, NGC~448 \citep{Katkov2016MNRAS.461.2068K}, PGC~66551 \citep{Katkov2024ApJ...962...27K}, and the sample of CR galaxies from the MaNGA survey presented by \citet{Gasymov2025ApJS..281...19G}.
In the latter work, the stellar mass of the CR disk in NGC~5717 was already estimated, but here we aim to refine that value.
For this purpose, we recover the stellar line-of-sight velocity distribution (LOSVD) non-parametrically and decompose the observed spectra into two stellar components with distinct kinematics and stellar population parameters.

The first necessary step in this analysis is a single-component full-spectrum fit using simple stellar population (SSP) models.
We used the full-spectrum fitting package \nb\ \citep{nburst_a, nburst_b}, which models the entire galaxy spectrum by extracting the information encoded in the absorption-line spectrum. 
We employed a grid of SSP spectra from the E-MILES models \citep{Vazdekis2015MNRAS.449.1177V,Vazdekis2016MNRAS.463.3409V}, computed with BaSTI isochrones \citep{Pietrinferni2004ApJ...612..168P} and a Kroupa Universal IMF \citep{Kroupa2001MNRAS.322..231K}. 
We used the base models, for which the stellar abundance pattern follows that of the Milky Way.
The model spectrum was broadened by convolution with a Gaussian stellar LOSVD.
To account for discrepancies between the model and observed spectra caused by imperfect calibration and dust attenuation, we applied a multiplicative polynomial continuum represented by 19th-degree Legendre polynomials.
The model also includes a set of strong emission lines (\Hg, \Hb, [O\iii], [O\oi], [N\ii], \Ha, [S\ii], etc.) represented by a common parametric Gaussian LOSVD profile independent of the stellar kinematics.
These emission lines were treated as additive components, and their fluxes were determined by solving a constrained linear problem at each step of the non-linear minimization loop. 
Both the emission-line templates and SSP models were pre-convolved with the wavelength-dependent instrumental resolution prior to the main minimization process.

The spectra were spatially binned to increase the signal-to-noise ratio of the stellar continuum.
We used the Voronoi binning technique \citep{Cappellari2003MNRAS.342..345C}, defining the signal as the median flux within the rest-frame wavelength range 5075--5125\,\AA\ and the noise as the median flux uncertainty over the same interval.
The final parameter maps are shown in Fig.~\ref{fig:kin_maps} and the result is consistent with the MaNGA Data Analysis Pipeline (DAP; \citealt{Westfall2019AJ....158..231W}; \citealt{Belfiore2019AJ....158..160B}).
The signatures of the CR disk, namely the $2\sigma$ peaks in the stellar velocity dispersion map, the ``S''-shaped rotation pattern in the stellar velocity map, and the gas-stellar counter-rotation, are all reproduced.

\begin{figure}
\centering
\includegraphics[width=0.49\textwidth]{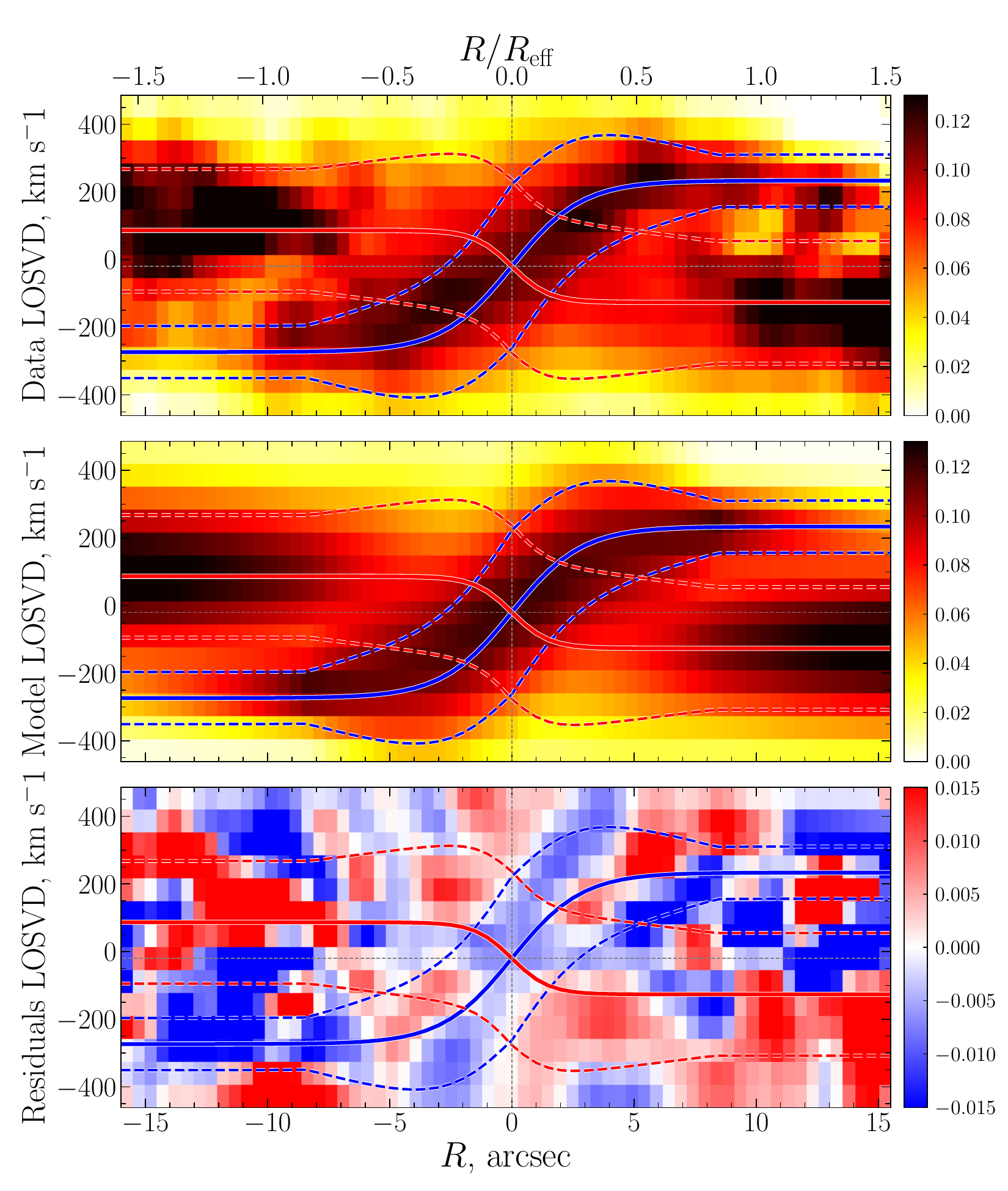}
\caption{Recovered non-parametric stellar LOSVD and its decomposition.
The plots, from top to bottom, show the extracted non-parametric stellar LOSVD along the major axis, the best-fitting model, and the residuals.
The color represents the normalized LOSVD amplitude, i.e. the relative contribution of stars to a given line-of-sight velocity bin at each spatial position along the slit.
The red and blue solid curves indicate the rotation velocities of the main (elliptical) and CR disk components, respectively; the dashed curves are offset from the corresponding rotation curves by one stellar velocity dispersion, $\sigma_\star$.
\label{fig:losvd_1d}}
\end{figure}

During the single-component full-spectrum fitting, the stellar LOSVD was approximated by a single Gaussian, which is formally inadequate for systems with kinematically decoupled stellar components, in particular galaxies hosting counter-rotating disks.
Therefore, to obtain physically meaningful kinematics, we recovered the stellar LOSVD non-parametrically by solving the linear inverse problem with regularization (for details of the method, see \citealt{Gasymov2024ASPC..535..279G}).
The recovery was performed independently in each Voronoi bin, producing a three-dimensional kinematic description of the galaxy in the $({\rm R.A.}, {\rm Dec.}, V)$ space.
We then reduced its dimensionality by extracting a position--velocity representation along a 1~arcsec pseudo-slit \citep[\,\textsc{pvextractor}\footnote{\url{https://pvextractor.readthedocs.io/}}]{Ginsburg2016ascl.soft08010G} oriented along the major axis of the CR disk (PA $\approx 24^\circ$).
As expected for a superposition of two CR components, the extracted LOSVD exhibits a pronounced X-shaped morphology (Fig.~\ref{fig:losvd_1d}), which we modeled as the sum of two components with exponential radial light profiles.
The best-fitting model yields a ``main'' component with a rotation amplitude of $107 \pm 3$ km~s$^{-1}$ and a constant velocity dispersion of $\sim 181 \pm 3$ km~s$^{-1}$ at large radii, and a dynamically colder counter-rotating disk with a rotation amplitude of $V_{\rm CR} = 254 \pm 2$ km~s$^{-1}$ and $\sigma_{\rm CR} = 77 \pm 2$ km~s$^{-1}$.
The parameter uncertainties were estimated from the distributions recovered in 10,000 Monte Carlo realizations.
From the fitted luminosity profiles, we infer an exponential scale length of $3.28 \pm 0.03$ arcsec ($h_{\rm CR} = 2.43 \pm 0.02$ kpc) for the CR disk and its contribution to the total luminosity of $W_{\rm CR} = (29 \pm 1)\%$ of the galaxy light.
The details of the modeling technique are described by \citet{Gasymov2025ApJS..281...19G}.

\subsection{Gas properties}
\subsubsection{Ionized gas}
\label{subsec:ion_gas}

\begin{figure*}
\centering
\includegraphics[width=0.98\textwidth]{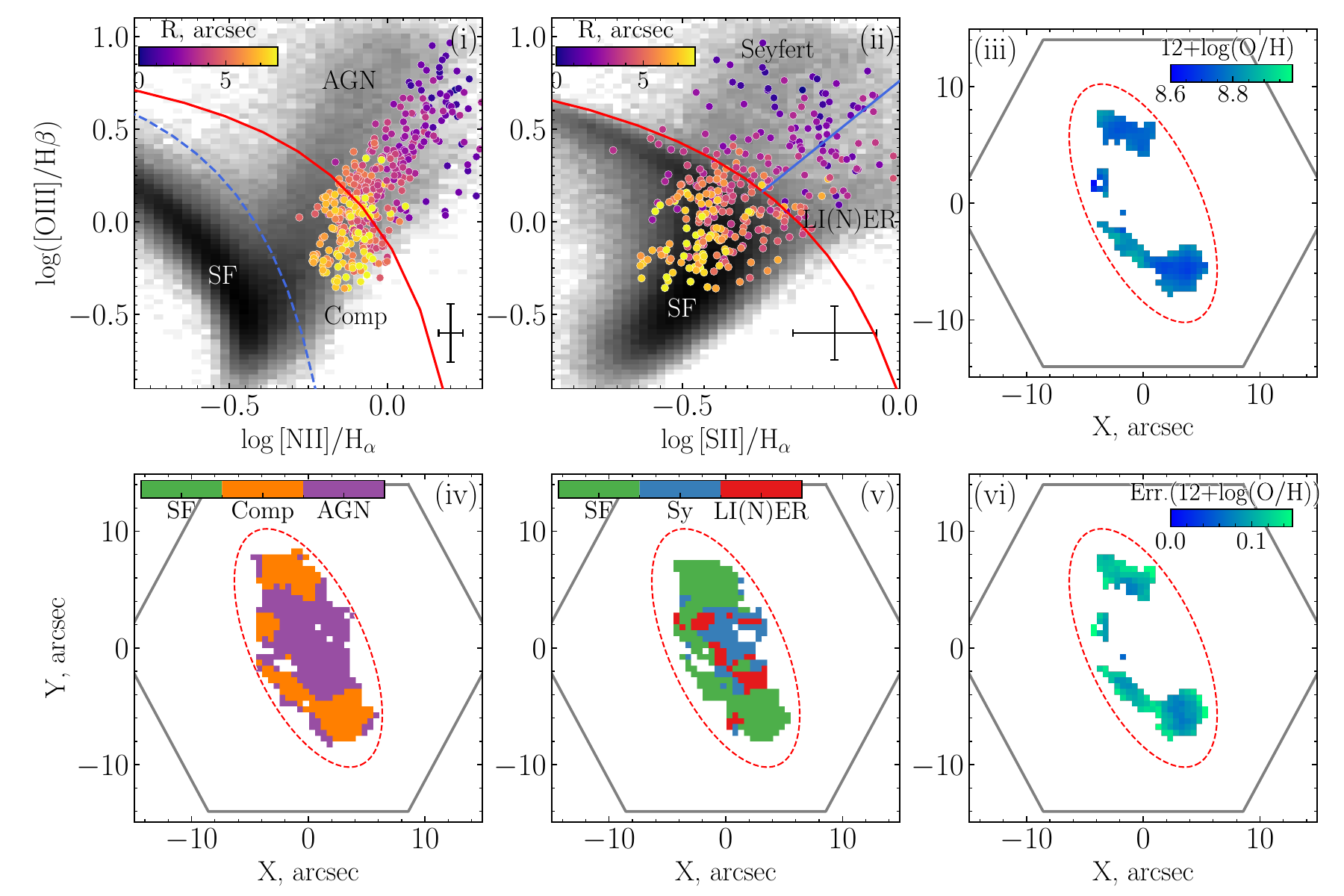}
\caption{BPT--[N\ii] and BPT--[S\ii] diagnostic diagrams (i, ii), gas-phase metallicity map with per-bin metallicity uncertainties (iii, vi), and the corresponding resolved BPT maps (iv, v).
The gray hexagons mark the MaNGA fields of view, and the red ellipses approximately trace the outer edge of the counter-rotating disk.
In the BPT--[N\ii] diagram (i), the blue dashed line shows the empirical demarcation between star-forming and composite regions \citep{kauffmann03}.
Other demarcation curves separate composite and AGN regions in the BPT--[N\ii] plane (i) and divide the BPT--[S\ii] diagram (ii) into star-forming, Seyfert, and LI(N)ER regions \citep{kewley01}.
The gray background shows the distribution of emission-line measurements from RCSED \citep[\url{rcsed.sai.msu.ru},][]{Chilingarian2017ApJS..228...14C}.
In the right column, the gas-phase metallicity (iii) and its uncertainty (vi) were computed using the S-calibration of \citet{Pilyugin2016MNRAS.457.3678P} for bins classified as composite in the BPT--[N\ii] and as star-forming in the BPT--[S\ii] diagrams.
\label{fig:bpt_met}}
\end{figure*}

The galaxy exhibits bright emission lines, but only within a limited central area that overlaps with the region of the CR disk (Fig.~\ref{fig:kin_maps}); emission lines are undetectable beyond $\sim1\,R_{\rm eff}$ ($\approx10$\,arcsec).
The ionized-gas velocity field closely follows the kinematics of the secondary stellar disk (Fig.~\ref{fig:kin_maps}), indicating a clear connection between the gaseous and stellar CR components.
In this section, we discuss the properties of the ionized gas in NGC~5717.

First, we identify the dominant ionization source of the ionized gas.
For this purpose, we used the Baldwin--Phillips--Terlevich (BPT) diagnostic diagrams \citep[BPT;][]{bpt}, adopting the demarcation lines of \citet{kewley01} and \citet{kauffmann03} to distinguish between star-forming (SF), composite (Comp), Seyfert (Sy), and active galactic nucleus (AGN) / low-ionization (nuclear) emission-line region (LI(N)ER) excitation.
The BPT classification (Fig.~\ref{fig:bpt_met}) indicates an AGN at the galaxy center, which is also supported by the presence of a compact radio source in the Very Large Array Sky Survey (VLASS) data \citep{Lacy2020PASP..132c5001L}.
The surrounding region is classified as ``composite'' in the BPT--[N\ii] and as ``star-forming'' in the BPT--[S\ii] diagrams.
We attribute this mismatch to the relatively large point spread function (PSF) of the MaNGA data (FWHM$_{\rm PSF}\simeq 2$--$2.5''$), which blends emission from the weakly star-forming disk with the AGN-dominated nucleus.
We note, however, that part of the extended low-ionization emission may also be associated with LI(N)ER-like excitation or shocks, which can occupy a similar region of the BPT diagrams \citep{Allen2008ApJS..178...20A}.
Under this assumption, we treat the intersection of the ``Comp.'' and ``SF'' regions (green and blue areas in the resolved BPT maps in Fig.~\ref{fig:bpt_met}) as dominated by star formation, and we use these spaxels (hereafter, ``SF bins'') in the analysis below.

The central disk is prominent in the Legacy RGB image (Fig.~\ref{fig:kin_maps}), but the unsharp-masked image as well as the SDSS imaging reveal significant dust content in the disk.
Accurate two-component spectral fitting therefore requires an estimate of the extinction affecting each stellar component.
For a disk embedded in a spheroidal component, we assume that the disk is attenuated by the full V-band extinction, $A_V$, whereas only half of the spheroidal light is affected (with the near-side half unattenuated and the far-side half attenuated by the full $A_V$).
To estimate $A_V$ in NGC~5717, we use the SF bins and derive the nebular extinction from the Balmer decrement, adopting the extinction curve of \citet{Fitzpatrick1999PASP..111...63F}.
The median observed decrement is (\Ha/\Hb)$_{\rm obs}=5.12$, which corresponds to $A_{V,{\rm gas}}\simeq 2.02$\,mag assuming a theoretical ratio (\Ha/\Hb)$_{\rm lab}=2.86$.
Applying the conversion of \citet{Calzetti2000ApJ...533..682C}, we obtain a stellar attenuation of $A_{V,{\rm st}}\simeq 0.89$\,mag.
The Galactic foreground extinction toward NGC~5717 is small, $A_{V,{\rm gal}}\simeq 0.04$\,mag \citep{Schlafly2011ApJ...737..103S}, and is neglected in the following analysis.

For our galactic chemical evolution modeling (Sect.~\ref{subsec:chem_mod}), an estimate of the gas-phase metallicity is essential because it traces the present-day chemical abundance of the gas.
For all SF bins, we derived oxygen abundances after correcting the emission-line fluxes for extinction, using the S-calibration of \citet{Pilyugin2016MNRAS.457.3678P}.
The resulting metallicity distribution and its per-bin uncertainties are shown in panels (iii) and (vi) of Fig.~\ref{fig:bpt_met}.
The metallicity shows no significant spatial variations; the median value is $12+\log({\rm O/H}) = (8.77 \pm 0.10)$~dex, where the uncertainty is the median measurement error.
This is slightly higher than the solar value, $(12+\log({\rm O/H}))_\odot = (8.69 \pm 0.04)$~dex \citep{Asplund2021A&A...653A.141A}.

\subsubsection{Neutral gas}
\label{subsec:HI_gas}

The presence of ionized gas does not necessarily imply the presence of a substantial neutral or cold star-forming gas reservoir.
A cross-match with the MaNGA-HI catalog returns an \HI\ detection for NGC~5717 with an estimated mass of $\log M_{\rm HI} / M_\odot \approx 9.24$.
However, the spectrum is noisy and the detection appears uncertain.
Moreover, the systemic velocity of the reported \HI\ detection, in the optical convention, is $(11{,}151 \pm 20)$~km~s$^{-1}$, which is offset by about 190 km~s$^{-1}$ from the corrected systemic velocity derived from the stellar kinematic map --- 10,963 km~s$^{-1}$.

A cross-match with the SDSS catalog within 4.5 arcmin of NGC~5717 ($\approx 200$ kpc; corresponding to FWHM$_{\rm HI}/2$) at the \HI\ systemic velocity reveals a small blue star-forming galaxy, PGC~2283962\footnote{\url{https://rcsed2.voxastro.org/data/galaxy/402755}}.
We therefore assume that the reported \HI\ detection is associated with this galaxy rather than with NGC~5717.
Following \citet{Stark2021MNRAS.503.1345S}, we estimate an upper limit on the \HI\ mass of NGC~5717 at its actual systemic velocity.
The upper-limit calculation requires an assumed H\,I linewidth, $W$ which we have estimated from the deprojected rotation amplitude of the CR disk.
The observed line-of-sight amplitude from the stellar kinematic model is $V_{\rm los}=247$\kms\ (Sect.~\ref{subsec:kin_fit}); correcting for the inclination of the outer ring, $i = \arccos(b/a) = \arccos(0.444) \approx 64^\circ$ (panel (ii) of Fig.~\ref{fig:kin_maps}), gives $V_{\rm rot} = V_{\rm los} / \sin i \approx 276$\kms.
We therefore adopt $W = 2V_{\rm rot} \approx 552$\kms, obtaining $\log (M_{\rm H\,I,lim}/M_\odot)\simeq9.13$.

This conclusion is also indirectly supported by the FASHI survey \citep{Zhang2024SCPMA..6719511Z}.
A cross-match with the group environment returns six H\,I sources, but none of them is located at the position of NGC~5717.
The minimum detected H\,I mass among these sources is $\log M_{H\,I}/ M_\odot\approx 9.31$, which is consistent with a detection limit of $\log M_{H\,I}/ M_\odot\sim 8.9$--9.1 at the group distance of $\sim160$ Mpc (see Fig.~3 in \citealt{Zhang2024SCPMA..6719511Z}).

\subsection{Stellar populations}
\label{subsec:2comp_fit}

To derive the stellar population properties of the disk and the elliptical component separately, we performed a two-component full-spectrum fit using the spectral decomposition technique \citep{Chilingarian2011MNRAS.412.1627C, Coccato2011MNRAS.412L.113C, Johnston2013MNRAS.428.1296J, Katkov2013ApJ...769..105K}. 
In this approach, the observed spectrum is represented as a sum of two E-MILES SSP templates from the same model grid as used in Sect.~\ref{subsec:kin_fit}, with independent Gaussian LOSVDs and free relative luminosity contributions, implemented within the \nb\ framework.
The decomposition is inherently sensitive to the spectral S/N, the instrumental resolution, and, most importantly, the line-of-sight velocity separation between the components. As a result, it becomes unreliable in the central region, where $\Delta V$ is small and the relative weights and stellar population parameters of the two components become highly degenerate.
A key difference from our previous applications of this method \citep{Katkov2016MNRAS.461.2068K, Katkov2024ApJ...962...27K, Gasymov2025ApJS..281...19G} is that here we allowed for different internal extinction in the two stellar components.
As indicated by the dust morphology (Fig.~\ref{fig:kin_maps}), the dusty structure is associated with the CR disk, while half of the surrounding elliptical component lies in front of the disk, and half lies behind it along the line of sight.
We therefore applied the full $A_V$ attenuation to the disk component (Sect.~\ref{subsec:ion_gas}), whereas the elliptical component was modeled as the sum of two stellar populations with identical kinematic parameters, ages, and metallicities: a far-side population attenuated by the full $A_V$ and a near-side population left unattenuated.
This prescription approximates a mixed line-of-sight geometry, in which roughly half of the elliptical light is located behind the inner dusty disk and half in front of it, thereby reducing systematic biases in the derived ages, metallicities, and relative component weights.

The spectra were binned into elliptical sectors aligned with the major axis of the disk (Fig.~\ref{fig:2comp_fit}).
We co-added spaxels within sectors centered on the galaxy nucleus and aligned with the disk's major axis (PA $\approx 24^\circ$), adopting an opening angle of $\pm20^\circ$ and sampling projected radii from 3.5 to 9.5 arcsec in steps of 1 arcsec.
The central region ($R < 3.5$ arcsec) was excluded because the two stellar components cannot be reliably separated there owing to their small kinematic offset and contamination from the AGN continuum.
This binning scheme yields a stellar-continuum S/N of $\simeq 50$ at small radii and $\simeq 30$ at the outer edge, allowing us to place robust constraints on the component flux contributions (weights) and SSP parameters. 
At radii comparable to the apparent edge of the disk, its contribution rapidly decreases and the spectra become dominated by the ``elliptical'' component; therefore, in these outer bins the spectra were fitted with only a single stellar component (Fig.~\ref{fig:2comp_fit}).
We note that co-adding spaxels with slightly different projected kinematics may mildly reduce the recovered rotation amplitudes and broaden the LOSVDs.
However, given the narrow sector geometry and the relatively weak kinematic gradients within each bin, this effect is minor and does not affect the derived stellar population properties.

\begin{figure*}
\centering
\includegraphics[width=0.57\textwidth]{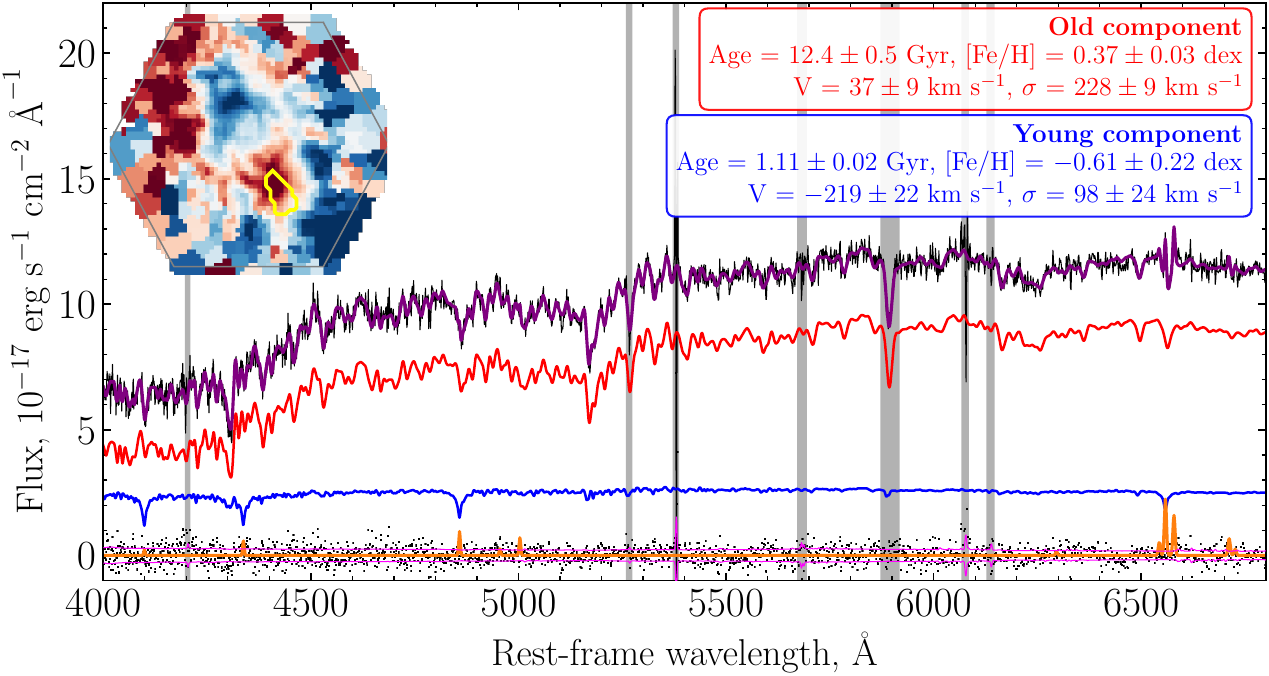}
\includegraphics[width=0.42\textwidth]{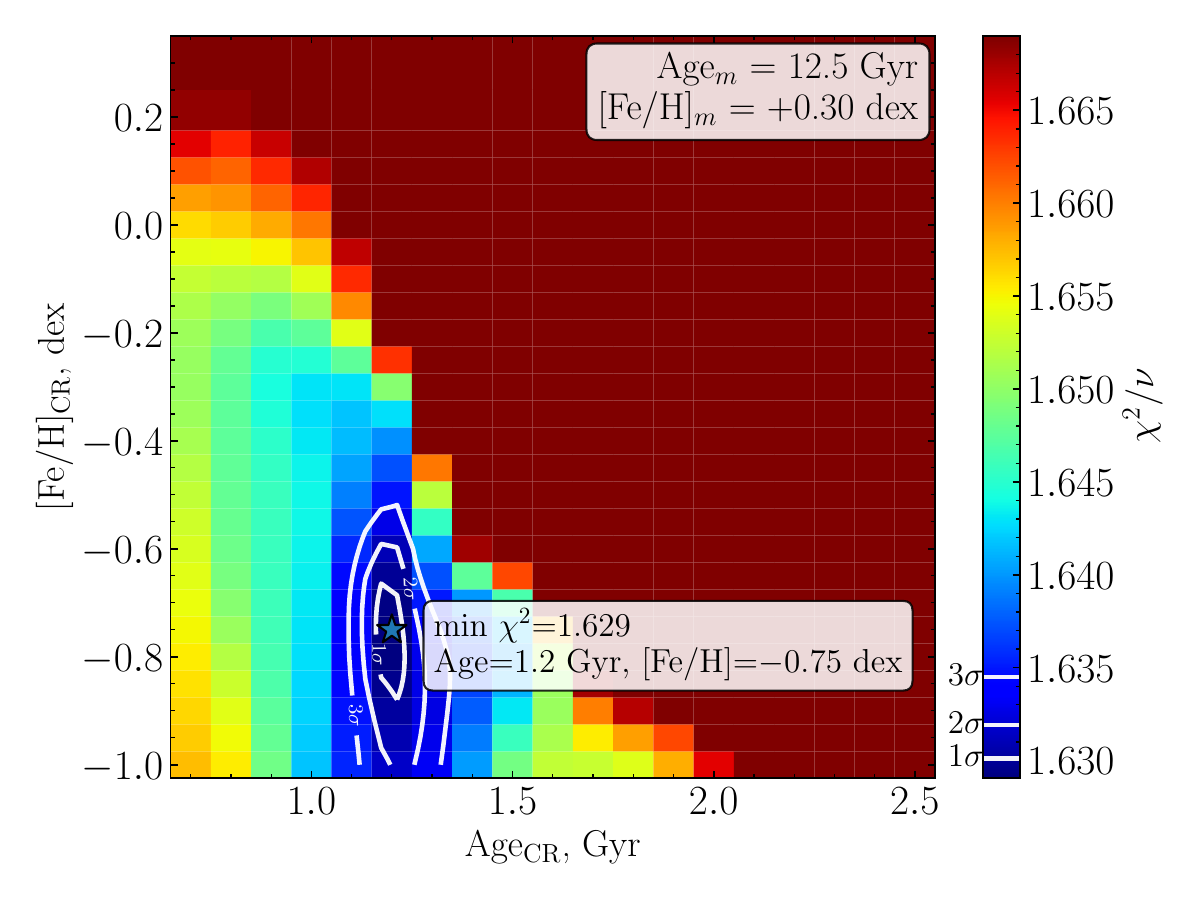}
\caption{\textbf{Left:} Two-component spectral decomposition.
The map in the upper-left corner shows the first-moment map of the recovered non-parametric stellar LOSVD (i.e., an analog of the stellar velocity map).
The MaNGA FoV is shown as a gray hexagon and the yellow region indicates the position of the used bin (see Sect.~\ref{subsec:2comp_fit}).
The black line represents the observed binned spectrum, while the thin magenta stripe shows the spectral uncertainties.
The purple line denotes the sum of the two stellar components and the emission lines (i.e., the best-fit model); the red line indicates the old ``main'' component, and the blue line the young CR disk.
The emission lines are shown in orange, while the gray dots indicate the residuals between the observed spectrum and the full model.
\textbf{Right:} $\chi^2$ scan for the stellar population parameters of the young component. 
The old stellar component was fixed at ${\rm Age}_m = 12.5$~Gyr and [Fe/H]$_m = +0.3$~dex. 
The color scale represents $\chi^2$ per degree of freedom, and the white contours indicate the 1, 2, and $3\sigma$ confidence levels. 
While a clear age--metallicity degeneracy is present for the young component, the allowed parameter space remains confined to low metallicities, [Fe/H] ${\rm_{CR}} < -0.5$ dex, and young ages, ${\rm Age_{CR}} < 1.4$~Gyr, at the $3\sigma$~level.
\label{fig:chi2_scan}}
\end{figure*}

\begin{figure*}
\centering
\includegraphics[width=\textwidth]{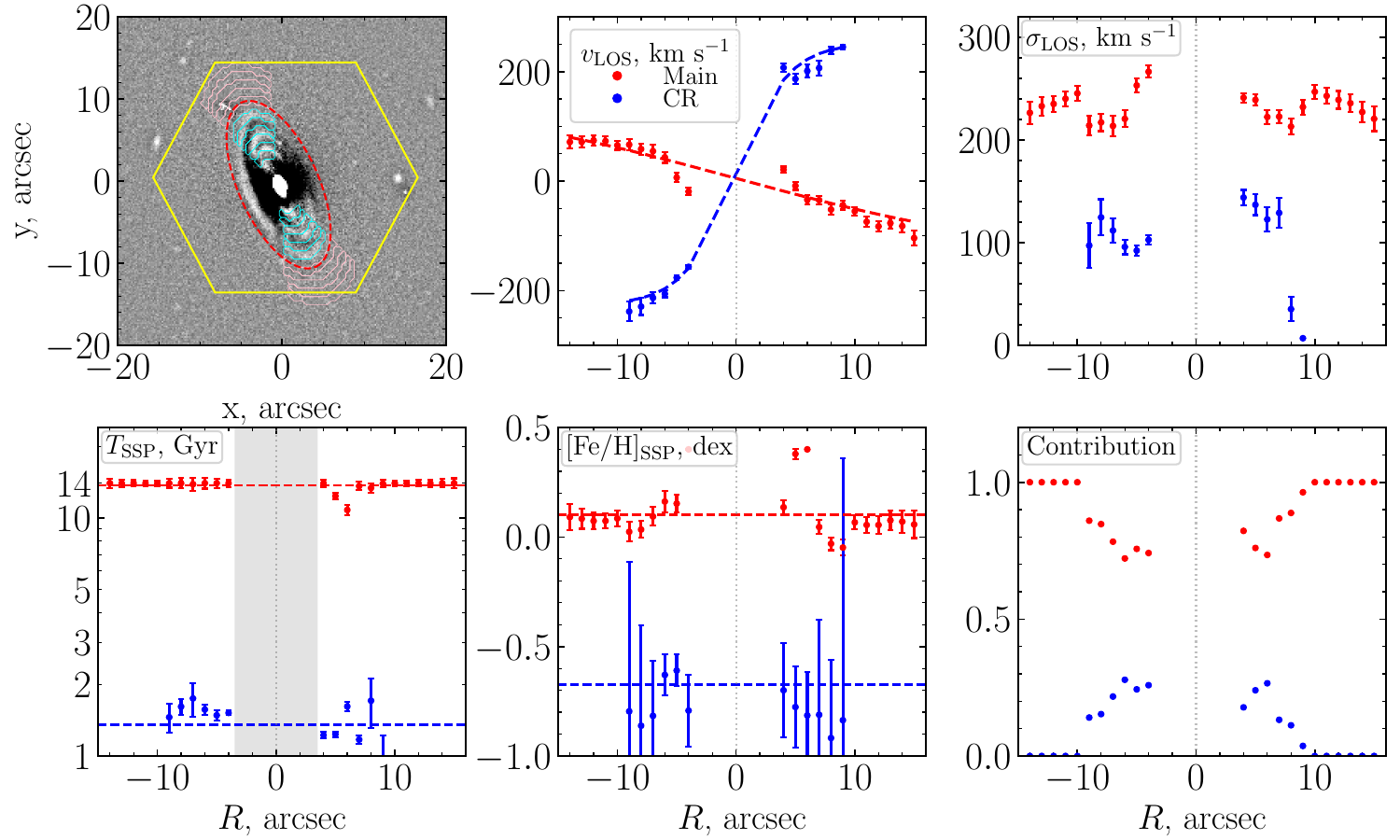}
\caption{Results of the two-component SSP spectral analysis.
The top-left panel shows the unsharp-masked CFHT $r$-band image of NGC~5717; the yellow hexagon indicates the MaNGA field of view, the red ellipse approximately traces the outer edge of the CR disk, and the cyan and pink elliptical sectors illustrate the spatial binning adopted for the ``inner'' two-component and ``outer'' one-component fits, respectively.
In the outer bins, where the CR component is not reliably detected, the spectra were fitted with a single (elliptical) component.
The central region was masked and excluded from the analysis.
Red and blue points correspond to the main (elliptical) and CR components, respectively.
From left to right and top to bottom, the remaining panels show: the line-of-sight stellar velocity $v_{\rm LOS}$ (with dashed curves showing the best-fitting $\tanh$ models), the line-of-sight stellar velocity dispersion $\sigma_{\rm LOS}$, the SSP-equivalent age $T_{\rm SSP}$ and metallicity ${\rm [Fe/H]}_{\rm SSP}$ (dashed lines indicate the median values with their estimated uncertainties), and the relative contribution of each component to the total spectrum at $5100\pm50$\,\AA\ (weights).
\label{fig:2comp_fit}}
\end{figure*}

The left panel of Fig.~\ref{fig:chi2_scan} shows an example of two-component spectral decomposition, including the observed spectrum, the two stellar components with an emission-line model, and the residuals.
Figure~\ref{fig:2comp_fit} then shows the radial behavior of the derived parameters.
The stellar population parameters of the components do not show strong radial gradients.
The kinematically ``hot'' elliptical component is old (Age$_{\rm ell} = 13.7~\pm~1$~Gyr, effectively an upper limit) and metal-rich ([Fe/H]$_{\rm ell} = +0.1~\pm~0.1$~dex), whereas the kinematically ``cold'' CR disk is young (Age$_{\rm CR} = 1.35~\pm~0.16$~Gyr) and metal-poor ([Fe/H]$_{\rm CR} = -0.67~\pm~0.09$~dex).
These values are weighted means over the radial bins shown in Fig.~\ref{fig:2comp_fit}. 
The uncertainties correspond to the weighted scatter of the measurements, after rescaling the fitting errors in each bin by $\sqrt{\chi_\nu^2}$.
The contribution of the CR disk also behaves as expected, steadily decreasing with radius.

To verify the derived stellar population parameters of the CR disk, we performed two additional tests using a large merged bin on the right side of the galaxy.
This bin was constructed as an elliptical sector aligned with the disk's major axis (PA $\approx 24^\circ$), with an opening angle of $10^\circ$ and spanning the radial range from 4 to 9~arcsec.
The first test consisted of a $\chi^2$ scan in the four-dimensional space of age and [Fe/H] for both stellar components, in order to reveal possible parameter degeneracies.
In this scan, the elliptical component consistently converges toward the old, metal-rich boundary of the model grid, consistent with the two-component fit.
In the right panel of Fig.~\ref{fig:chi2_scan}, we show the $\chi^2$ projection for the CR disk at the best-fitting grid point of the old stellar population.
Although the confidence contours are elongated along the metallicity axis, the CR disk remains confined to young ages and low metallicities within the $3\sigma$ confidence region.
The second test was motivated by the possible dependence of the fit on the initial guesses for the stellar population parameters.
However, fits performed both with free component weights and with fixed component contributions demonstrate that the described minimum is global, although several local minima with higher $\chi^2$ values are also present.

The corresponding $r$-band stellar mass-to-light ratios of the components are $(M/L)_{\rm ell} = (3.54~\pm~0.57)\,M_\odot/L_\odot$ and $(M/L)_{\rm CR}=(0.44~\pm~0.03)\,M_\odot/L_\odot$.
From the LOSVD model of the galaxy (see Sect.~\ref{subsec:kin_fit}), we estimated the total optical-light contribution of the CR disk to be $W_{\rm CR} = (29 \pm 1)\%$.
After applying the corresponding mass-to-light ratio correction, the total mass fraction of the CR disk becomes $W_{\rm m,CR} = (5 \pm 1)\%$, implying a stellar mass of $\log(M_{\star,\rm CR}/M_\odot) = 9.9 \pm 0.1$.
We assume that the remaining stellar mass of the galaxy is associated with the elliptical component, whose mass is therefore $\log(M_{\star,\rm ell}/M_\odot) = 11.18 \pm 0.01$.

\begin{table}[ht]
\centering
\caption{Properties of the two stellar components and the gaseous component of the galaxy.
$V_{\rm rot}$ and $\sigma_{\rm comp}$ denote the rotation velocity and the outer constant velocity dispersion of the corresponding stellar component, respectively (Sect.~\ref{subsec:kin_fit}).
$T_{\rm SSP}$ and $[\mathrm{Z}/\mathrm{H}]_{\rm SSP}$ are the SSP-equivalent age and metallicity, $M/L$ is the stellar mass-to-light ratio, and $M{\rm comp}$ is the stellar mass of the corresponding component (Sect.~\ref{subsec:2comp_fit}).
$A_{V,{\rm st}}$ is the V-band stellar extinction in the CR disk, and $12+\log({\rm O/H})$ is the gas-phase metallicity of the ionised gas (Sect.~\ref{subsec:ion_gas}).
The upper limit on the \HI\ mass, $\log(M_{\mathrm{H\,I}}/M_\odot)$, is derived from the sensitivity limits of the available surveys (Sect.~\ref{subsec:HI_gas}).
}
\label{tab:spectrophotometric}
\renewcommand{\arraystretch}{1.15}
\setlength{\tabcolsep}{8pt}

\begin{tabular}{lcc}
\hline\hline
Parameter & Elliptical comp. & CR disk \\
\hline

$V_{rot}$, \kms\ & $107 \pm 3$ & $254 \pm 2$ \\
$\sigma_{comp}$, \kms\ & $181 \pm 3$ & $77 \pm 2$ \\
$T_{\rm SSP}$, Gyr & $13.7 \pm 1$ & $1.35 \pm 0.16$ \\
$[\mathrm{Z}/\mathrm{H}]_{\rm SSP}$, dex & $+(0.1 \pm 0.1)$ & $-(0.67 \pm 0.09)$ \\
$M/L$, $M_\odot/L_\odot$ & $3.54 \pm 0.57$ & $0.44 \pm 0.03$ \\
$M_{\mathrm{comp}}$, $10^9\,M_\odot$ & $8 \pm 2$ & $151 \pm 4$ \\

\hline 

$A_{V,{\rm st}}$ & \multicolumn{2}{c}{$0.89$ mag} \\
$(12+\log{\rm O/H})$, dex & \multicolumn{2}{c}{$8.77 \pm 0.10$} \\
$\log M_{\mathrm{H\,I}}/M_\odot$ & \multicolumn{2}{c}{$\le 9.13$} \\

\hline
\end{tabular}
\end{table}

\subsection{Properties of the galaxy group}

NGC~5717 occupies an unusual environmental niche among galaxies hosting CR disks.
A cross-match with the galaxy group catalog of \citet{Yang2007ApJ...671..153Y} shows that it belongs to a galaxy group and, moreover, it is the brightest galaxy in the group, a ``brightest group galaxy" (BGG).
NGC~5717 is the first identified BGG hosting a counter-rotating stellar disk, whilst galaxies with kinematically decoupled stellar components are generally found in relatively sparse environments \citep{Gasymov2025ApJS..281...19G}.

\begin{figure}
\centering
\includegraphics[width=0.48\textwidth]{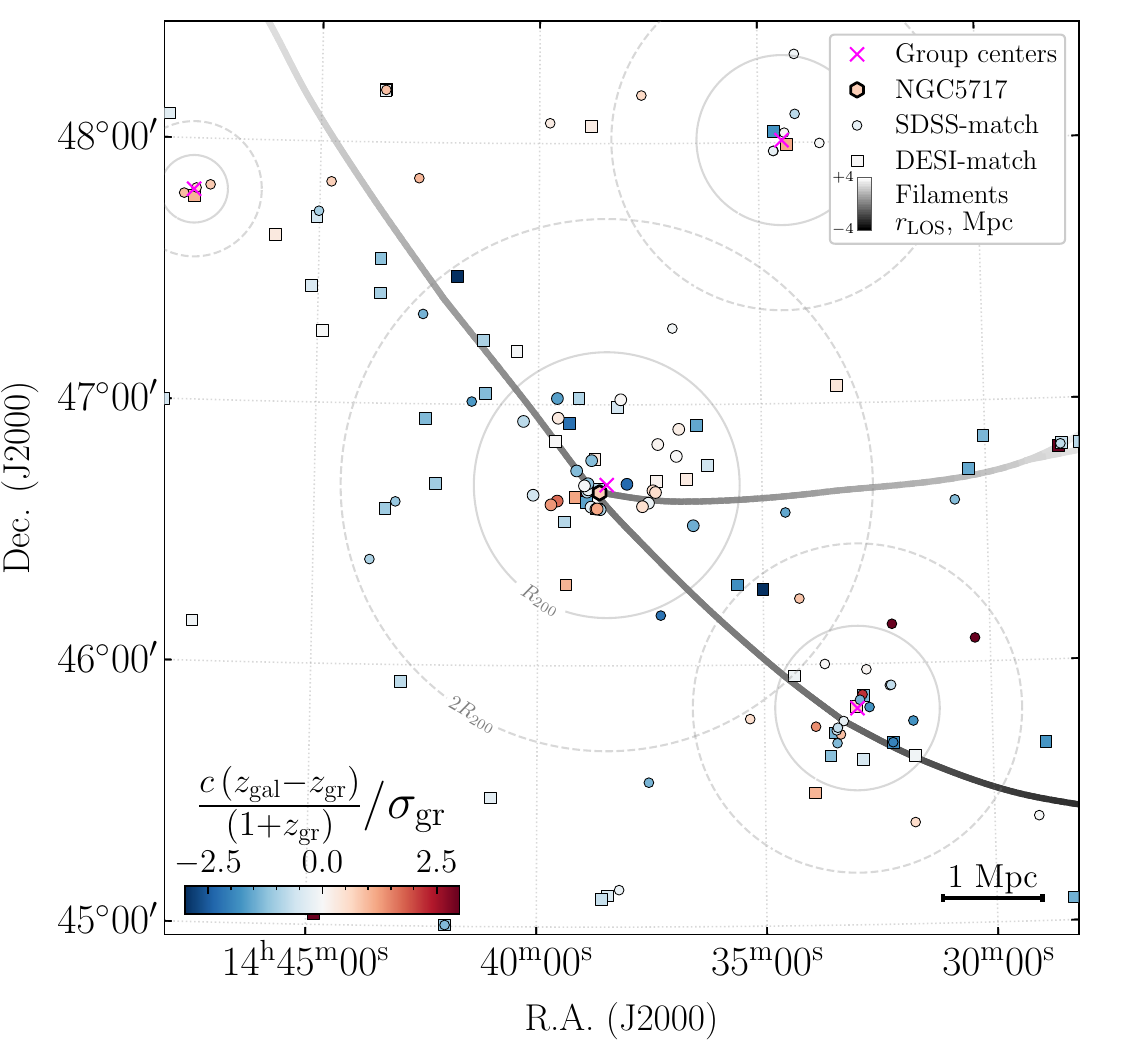}
\includegraphics[width=0.48\textwidth]{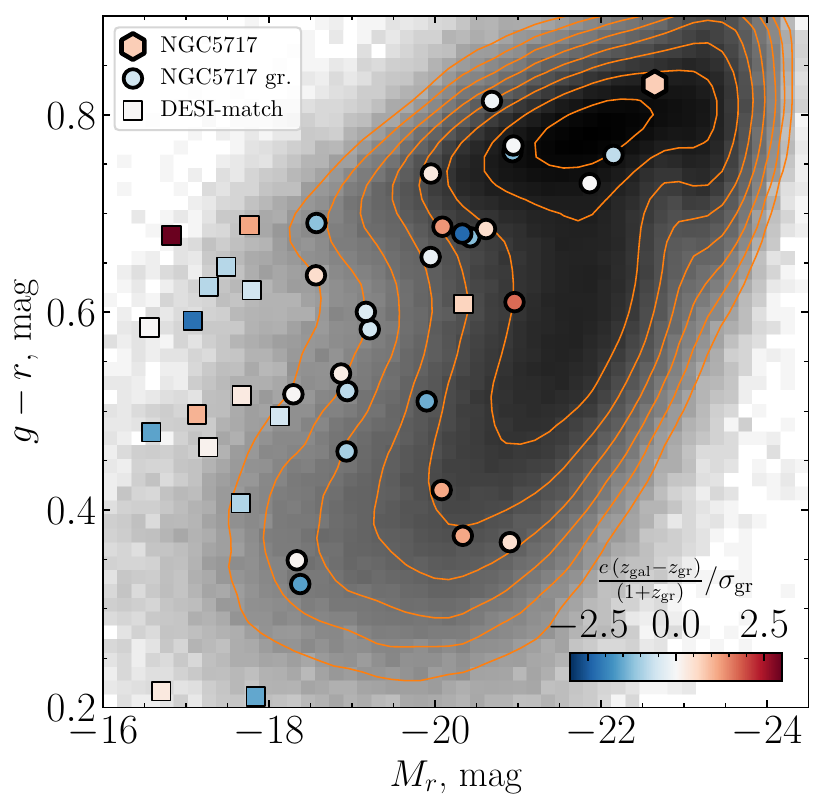}
\caption{\textbf{Top:} Spatial distribution of galaxies matched in SDSS and DESI. 
The cross-match was performed within a $1.7^\circ$ radius and $\pm1000$\,km\,s$^{-1}$ of the group center and systemic velocity. 
The group center is marked by a magenta cross; the hexagon denotes NGC~5717. 
Circles and squares indicate galaxies from the SDSS and DESI matches, respectively. 
Points are color-coded by the line-of-sight velocity offset from the group's systemic velocity ($10{,}908$\,km\,s$^{-1}$), normalized by the group's velocity dispersion ($209$\,km\,s$^{-1}$; Table~\ref{tab1}). 
The solid and dashed gray circles indicate $R_{200}$ and $2R_{200}$ for the groups matched in the environment of NGC~5717. 
The gray lines show the projection of the cosmic-web filaments \citep{Kraljic2020MNRAS.491.4294K} onto the R.A.--Dec. plane.
Their grayscale encodes r$_{\rm LOS}$ relative to the group center, ranging from $-4$ to $+4$~Mpc, from black to white.
\textbf{Bottom:} Color--magnitude diagram (CMD) for the matched NGC~5717 group members within $R_{200}$. 
Photometry was taken from the SDSS catalog \citep{Abdurro'uf2022ApJS..259...35A} and corrected with $K$-corrections \citep{Chilingarian2010MNRAS.405.1409C}.
The gray background shows the magnitude distribution from RCSED, and the orange curves indicate iso-density contours of this distribution \citep[\url{rcsed.sai.msu.ru},][]{Chilingarian2017ApJS..228...14C}.
\label{fig:group_plots}}
\end{figure}

The group catalog of \citet{Yang2007ApJ...671..153Y}, based on SDSS DR4 data, identifies 27 members in the group.
Of these, 24 have SDSS-based redshift measurements, while three galaxies, including NGC~5717, were obtained from external spectroscopic catalogs \citep{Yang2007ApJ...671..153Y}.
To improve the census of the environment, we cross-matched the system with DESI DR1 \citep{DESICollaboration2026AJ....171..285D} and searched for galaxies within $R_{200}$ and within $\pm3\sigma_{\rm cl}$ of the group systemic velocity (Table~\ref{tab1}), using projected coordinates and redshift, which revealed 16 additional candidate members.
The environment of NGC~5717, including the cross-matched galaxies, is shown in the top panel of Fig.~\ref{fig:group_plots}; in addition to the main system, three smaller nearby groups are visible.
The galaxies associated with the group within $R_{200}$ are also shown in the color--magnitude diagram in the bottom panel of Fig.~\ref{fig:group_plots}.
The majority of galaxies are fainter than $M_r = -20$, and they distribute along the red sequence and in the blue cloud, with approximately equal numbers in each.
The small magnitude gaps between the brightest group members, $\Delta M_{12}=0.50$ mag and $\Delta M_{14}=1.69$ mag, are well below the thresholds of 2 and 2.5 mag, respectively, indicating that this system is not a fossil group \citep{Dariush2010MNRAS.405.1873D}.

The dynamical state of the group is not entirely clear.
Its elongated projected structure (see Fig.~\ref{fig:group_plots}) and the offset of the BGG from the spatial and kinematic centers of the group ($\Delta R \approx 130$~kpc and $\Delta V \approx 147$~\kms) may indicate that the system has not yet fully relaxed \citep{Raouf2014MNRAS.442.1578R,DePropris2021MNRAS.500..310D}.
On the other hand, the galaxy velocity distribution is close to Gaussian, which is consistent with a dynamically relaxed system.
In the absence of sufficiently deep X-ray observations that could provide an independent constraint on the dynamical state, we therefore conservatively treat the group as virialized.
For a group of this halo mass, reaching such a state may require a few Gyr \citep{Poole2016MNRAS.459.3025P}.

We also examined the cosmic-web filaments in the environment of NGC~5717; hereafter, we refer to these as ``cosmic-web filaments'' to distinguish them from cold-gas ``cosmic filaments''. 
We used a DisPerSE cosmic-web skeleton constructed from the galaxy group catalog of \citet{Wang2014MNRAS.439..611W}, following the methodology of \citet{Kraljic2020MNRAS.491.4294K}. 
By matching NGC~5717 to the galaxy catalog corrected for the fingers-of-god effect and used to construct the cosmic web, we found that it lies only about $170$~kpc from the nearest filament branch.
Using the three-dimensional catalog position of the galaxy, we projected the surrounding cosmic-web structure onto the sky plane (top panel of Fig.~\ref{fig:group_plots}).
The sky plane coordinates were calculated as
\begin{equation}
\begin{aligned}
\mathrm{R.A.} &= \operatorname{atan2}(y,x),\\
\mathrm{Dec.} &= \arcsin\left(z~/ \sqrt{x^2+y^2+z^2}\right),
\end{aligned}
\end{equation}
where $x$, $y$, and $z$ are the Cartesian coordinates of the segment endpoints.
We identify two cosmic-web filaments that cross the NGC~5717 group and intersect close to its center (Fig.~\ref{fig:group_plots}).
The longer filament, extending from the upper-left to the lower-right corner of the figure, is inclined by about $24^\circ$ to the sky plane when averaged over the local volume, whereas the second filament is inclined by about $32^\circ$.

\subsection{Chemical evolution model}
\label{subsec:chem_mod}

\subsubsection{Framework}

To interpret the derived stellar population parameters and gas properties, we constructed a model for the chemical enrichment history.
The two-component stellar population fitting (see Fig.~\ref{fig:2comp_fit}) and the ionized-gas analysis (see Fig.~\ref{fig:bpt_met}) do not reveal significant radial gradients or strong spatial variations in the relevant properties.
This suggests that the secondary component of the galaxy can be approximated as a single well-mixed zone with a uniform star formation history and metallicity.
Under this assumption, the system can be described with a simple chemical evolution model in the instantaneous recycling approximation \citep[IRA;][]{Schmidt1963ApJ...137..758S, Pagel1975MNRAS.172...13P, Tinsley1980FCPh....5..287T}.

This approximation assumes that stars with masses $m < 1\,M_\odot$ have effectively infinite lifetimes, whereas more massive stars die instantaneously and enrich the interstellar medium (ISM) with newly synthesized metals, which are assumed to mix with the gas immediately.
As a result, at any given moment only a fraction $R$ of the stellar mass is returned to the ISM.
The total mass fractions of metals and oxygen are $y_Z$ and $y_O$, respectively, and are synthesized by massive stars and released into the ISM (the so-called yield per stellar generation).
These parameters depend primarily on the adopted initial mass function (IMF) and weakly on metallicity \citep{Vincenzo2016MNRAS.455.4183V}.
We adopted the values from \citet{Spitoni2017A&A...599A...6S} for the \citet{Kroupa2001MNRAS.322..231K} IMF: $R = 0.441$, $y_Z = 0.0631$, and $y_O = 0.0407$.

We constructed a model within the IRA framework as described by \citet{Spitoni2017A&A...599A...6S}.
The simplest physically motivated description of an embedded disk evolving inside a massive elliptical galaxy would be a classical \textit{closed-box} model. 
However, as we show below, the low metallicity of the CR disk cannot be reproduced within a closed-box framework, requiring efficient outflows to remove the excess metals produced by stellar evolution.

We therefore consider a more general chemical evolution model that includes gas infall $I(t)$, and gas outflow $O(t)$, and depends on the initial mass function (IMF) and the star formation rate (SFR) $\psi(t)$.
In our model, we assume the \citet{Kroupa2001MNRAS.322..231K} IMF, which was also used to construct the stellar libraries employed in the full-spectrum fitting (Sect.~\ref{subsec:2comp_fit}).
\citet{Spitoni2017A&A...599A...6S} formulate the SFR as directly proportional to the gas mass, i.e., following the linear Schmidt law \citep{Schmidt1959ApJ...129..243S}:
\begin{equation}
    \psi(t) = S \times M_{\rm gas}(t),
\end{equation}
where $S$ is the star formation efficiency (SFE, the reciprocal of the gas depletion time), and $M_{\rm gas}$ is the gas mass of the galaxy at time $t$.
The gas infall rate is described by an exponentially decaying law \citep{Chiosi1980A&A....83..206C}:
\begin{equation}
    I(t) = A e^{-t/\tau},
\end{equation}
where $A$ is the infall amplitude and $\tau$ is the infall timescale.
The gas outflow is assumed to scale linearly with the star formation rate \citep{Matteucci1983A&A...123..121M}, and we assume that the removed gas does not return to the galaxy:
\begin{equation}
    O(t) = \lambda \psi(t),
\end{equation}
where $\lambda$ is the mass-loading factor, which characterizes the efficiency of gas removal by galactic winds.

Combining these prescriptions, the evolution of the total mass, gas mass, and metal mass in the galaxy, $M_{\rm tot}(t)$, $M_{\rm gas}(t)$, and $M_{\rm Z}(t)$, is described by the following system of differential equations:
\begin{equation}
 \begin{cases}
   \dot{M}_\mathrm{tot}(t) = A e^{-t/\tau} - \lambda \psi(t);
   \\
   \dot{M}_\mathrm{gas}(t) = - (1 - R) \psi(t) + A e^{-t/\tau} - \lambda \psi(t);
   \\
   \dot{M}_\mathrm{Z}(t) = (-Z(t) + y_Z) (1 - R) \psi(t) + Z_\mathrm{inf} A e^{-t/\tau} -  \lambda Z_\mathrm{out} \psi(t),
 \end{cases}
 \label{eq:chem_evol_equation_set}
\end{equation}
where $Z_\mathrm{inf}$ and $Z_\mathrm{out}$ are the metallicities of the inflowing and outflowing gas, respectively.
We assume a constant inflow metallicity $Z_\mathrm{inf}$.

Our definition of the outflow metallicity differs slightly from that adopted by \citet{Spitoni2017A&A...599A...6S}.
Previous studies have shown \citep{Dalcanton2007ApJ...658..941D, Katkov2024ApJ...962...27K} that the outflowing gas is expected to be more metal-rich than the ISM with metallicity $Z(t)$.
We therefore adopt a prescription in which the outflow is enriched by supernova ejecta:
\begin{equation}
    Z_\mathrm{out}(t) = \epsilon Z(t) + (1 - \epsilon)Z_\mathrm{SN},
\end{equation}
where $\epsilon$ controls the relative contribution of the ISM and SN ejecta to the outflow: $\epsilon = 0$ corresponds to an outflow consisting entirely of SN ejecta, whereas $\epsilon = 1$ corresponds to an outflow with the ISM metallicity.
The metallicity of the supernova ejecta is $Z_\mathrm{SN}=\eta y_Z$, where $\eta=4.5-5.2$ for the \citet{Kroupa2001MNRAS.322..231K} IMF \citep{Dalcanton2007ApJ...658..941D}.
In our model, we adopt $\eta=4.5$, although the final results do not depend strongly on this choice.

The stellar mass is computed as the difference between the total baryonic mass and the gas mass of the galaxy: $M_\star (t) = M_\mathrm{tot}(t) - M_\mathrm{gas}(t)$.
The ISM metallicity $Z(t)$ is defined as the ratio of the metal mass to the gas mass: $Z(t) = M_\mathrm{Z}(t) / M_\mathrm{gas}(t)$.
Recalling that $M_\mathrm{Z}(t) = Z(t) M_\mathrm{gas}(t)$ and differentiating this expression with respect to time, we obtain the following differential equation for the metallicity:
\begin{equation}
    \dot{Z} (t) = (1-R) y_Z S - \lambda (1 - \epsilon) (Z_\mathrm{SN} - Z(t)) S + \frac{(Z_\mathrm{inf} - Z(t)) A e^{-t/\tau}}{M_\mathrm{gas}(t)}.
    \label{eq:chem_evol_Zdot}
\end{equation}

\subsubsection{Application to the galaxy}
\label{subsec:chem_mod_app}

We now apply the model described above to NGC~5717.
We consider a scenario in which the CR disk formed \textit{in situ} from externally accreted gas (Sect.~\ref{subsec:origin}), while the formation of the elliptical component is not modeled explicitly, since it was most likely assembled through a major merger or a sequence of dry minor mergers rather than having formed directly from an initial gas reservoir \citep{DeLucia2007MNRAS.375....2D, Montenegro-Taborda2023MNRAS.521..800M}.

There are five observational constraints that characterize the CR disk and the present-day gas content:
\begin{itemize}
    \item Median gas-phase metallicity $(12+\log{\rm O/H}) = (8.77 \pm 0.10)$~dex (see Sect.~\ref{subsec:ion_gas});
    \item Upper limit on the \HI\ mass, $\log M_{\mathrm{H\,I}}/M_\odot \le 9.13$ (see Sect.~\ref{subsec:HI_gas});
    \item SSP-equivalent (luminosity-weighted) age of the stellar population, ${\rm Age} = (1.35\pm0.16)$~Gyr (see Sect.~\ref{subsec:2comp_fit});
    \item SSP-equivalent (luminosity-weighted) metallicity of the stellar population, [Z/H] $=(-0.67\pm0.09)$~dex (see Sect.~\ref{subsec:2comp_fit});
    \item Estimated stellar mass of the CR disk, $\log M_{\star}/M_\odot \approx 9.9$ (see Sect.~\ref{subsec:2comp_fit}).
\end{itemize}

The model depends on six free parameters, with the adopted range given in parentheses:
\begin{itemize}
    \item Star formation efficiency $S$ (fixed to $2.0$~Gyr$^{-1}$, i.e., $t_{\rm dep} =0.5$~Gyr, see discussion in Sect.~\ref{subsec:SF_feautures});
    \item Infall timescale $\tau$ ($0.01{\rm~Gyr} \le \tau \le 10{\rm~Gyr}$);
    \item Onset time of infall $T$ ($1.35{\rm~Gyr} \le T \le 4{\rm~Gyr}$);
    \item Infall metallicity $Z_{\rm inf}$ ($2\cdot10^{-5} \le Z_{\rm inf} \le 2\cdot10^{-2}$, corresponding to $-3 \le [Z_{\rm inf}/H] \le 0$);
    \item Mass-loading factor $\lambda$ ($0 \le \lambda \le 1$);
    \item Contribution of SN ejecta to the outflow, $\epsilon$ ($0 \le \epsilon \le 1$).
\end{itemize}

To simplify the model, the gas infall amplitude $A$ is not treated as a free parameter. 
Instead, at each step of the minimization it is computed so that the model reproduces the observed final stellar mass.

\begin{figure*}
\centering
\includegraphics[width=\textwidth]{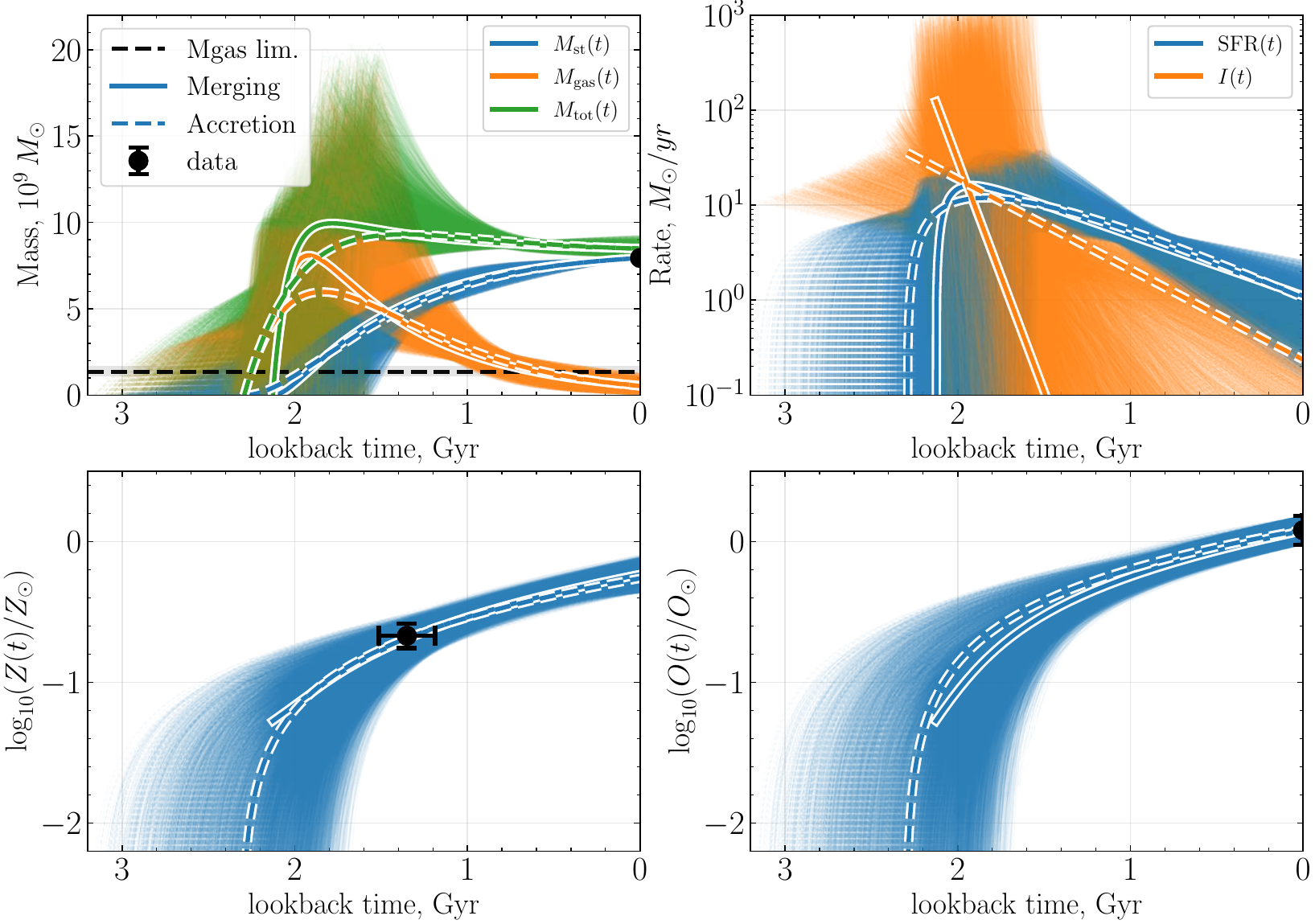}
\caption{Selected ``good'' chemical evolution models obtained from the random sampling of the parameter space.
The top-left panel shows the evolution of the stellar mass $M_{\rm st}(t)$ (blue), gas mass $M_{\rm gas}(t)$ (orange), and total baryonic mass $M_{\rm tot}(t)$ (green) as a function of lookback time.
The top-right panel presents the corresponding star formation rate $\psi(t)$ (blue) and infall rate $I(t)$ (orange).
The bottom panels show the evolution of the stellar metallicity, expressed as $\log_{10}(Z/Z_\odot)$ (left), and the gas-phase oxygen abundance, $\log({\rm O/H})/({\rm O/H})_\odot$ (right), relative to the solar value.
The black points indicate the observational data, while the dashed black line marks the present-day upper limit on the gas mass.
The two representative solutions discussed in Sect.~\ref{subsec:chem_mod_app} are highlighted by white solid and dashed curves.
The corresponding best-fitting parameter sets are marked in Fig.~\ref{fig:chem_corner_plot}.
\label{fig:chem_models}}
\end{figure*}

\begin{figure*}
\centering
\includegraphics[width=\linewidth]{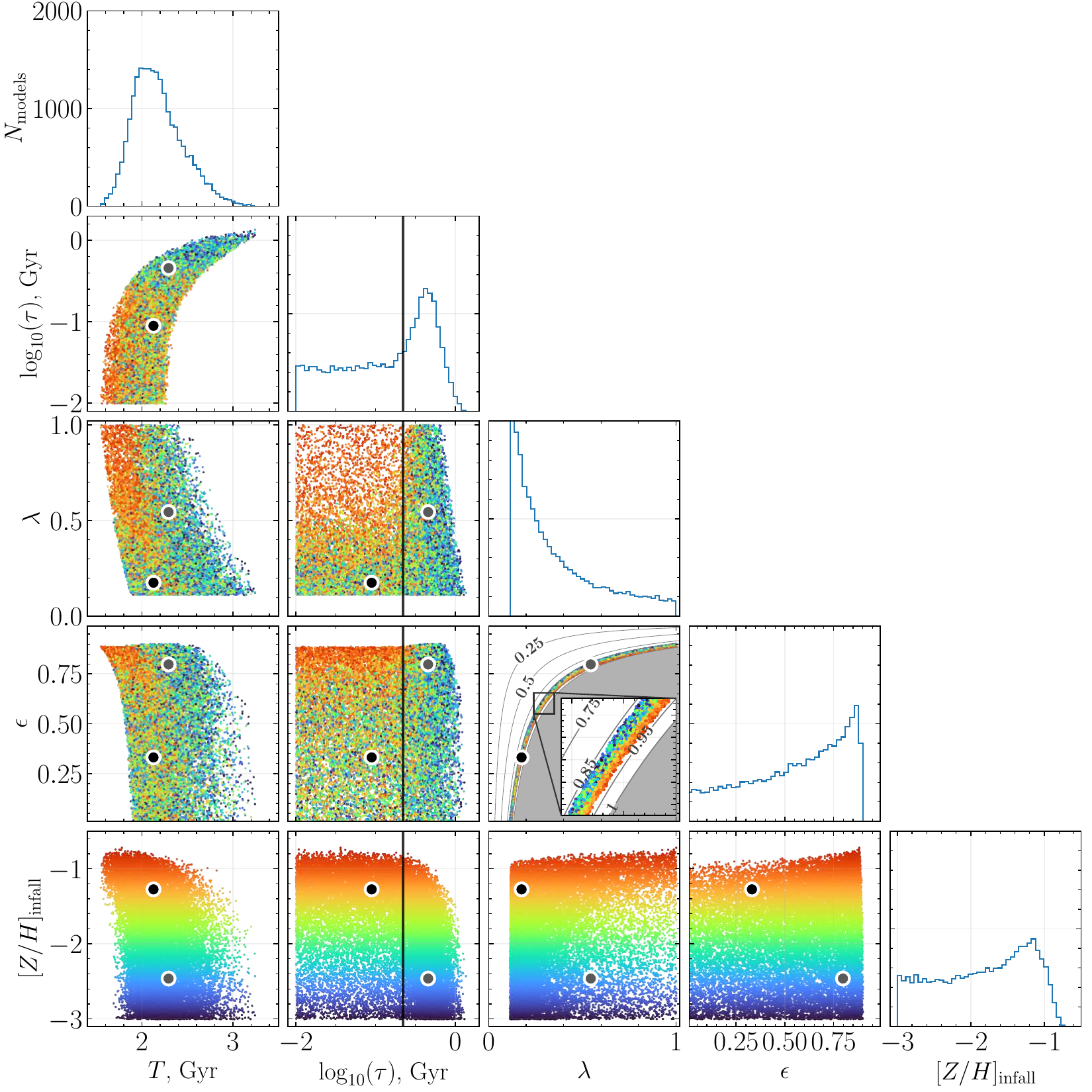}
\caption{Corner plot showing the parameter distributions of the ``good'' models presented in Fig.~\ref{fig:chem_models}.
The diagonal panels show the one-dimensional marginalized distributions of the model parameters, while the off-diagonal panels present their pairwise correlations.
Here, $T$ is the onset time of gas infall, $\tau$ is the infall timescale, $Z_{\rm inf}$ is the metallicity of the infalling gas, $\lambda$ is the mass-loading factor of the outflow, and $\epsilon$ controls the relative contribution of the ISM and SN ejecta to the outflow metallicity.
Two main degeneracies are apparent, namely between $T$ and $\tau$, and between $\lambda$ and $\epsilon$.
In the $\lambda$--$\epsilon$ plane, iso-contours of the metal-loss fraction $f_{\rm out}$ are overplotted for $Z = Z_{\rm CR}$ and correspond to $f_{\rm out} = 0.25$, 0.5, 0.75, 0.85, 0.95, and 1.
The gray-shaded region corresponds to $f_{\rm out} > 1$ and is non-physical.
The black and gray points mark the two representative minimum-$\chi^2$ solutions discussed in Sect.~\ref{subsec:chem_mod_app}, corresponding to the short-timescale metal-enriched infall and the longer-timescale metal-poor accretion scenarios, respectively.
The vertical black line indicates $\tau = 0.2$~Gyr.
\label{fig:chem_corner_plot}}
\end{figure*}

The model is overparameterized, and many combinations of parameters can reproduce the observed properties of the galaxy.
Rather than searching for a unique solution, we therefore aim to determine the region of parameter space consistent with the observational constraints.
To do this, we randomly sampled the model parameters within the ranges given above and retained those realizations that reproduce all observed quantities within their 1$\sigma$ uncertainties.
All such ``good'' models are shown in Fig.~\ref{fig:chem_models}, and the corresponding parameter distributions are presented in Fig.~\ref{fig:chem_corner_plot}.

At first, two main degeneracies are evident in the model parameter space: the $T$--$\tau$ degeneracy, which governs the time distribution of the accreted mass, and the $\lambda$--$\epsilon$--$Z_{\rm inf}$ degeneracy, which controls the balance between metal supply and metal loss in the galaxy.
The latter was already noted by \citet{Katkov2024ApJ...962...27K}, although it was not demonstrated there as explicitly as in the present case.
The $\lambda$--$\epsilon$ degeneracy reflects the chemical enrichment of the gas outflow, and to illustrate it we use the parameter $f_{\rm out}$ \citep{Tinsley1980FCPh....5..287T, Lian2018MNRAS.474.1143L}.
This parameter represents the fraction of metals removed by the outflow.
In our model, it depends on the ISM metallicity $Z$ and can be written as:
\begin{equation}
    \label{eq:f_out}
    f_{\rm out} = \dfrac{\lambda(\epsilon Z(t) + (1-\epsilon)\eta y_Z)}{RZ(t) + y_Z (1-R)}
\end{equation}
The parameter $f_{\rm out}$ must be less than 1, and this condition can be illustrated in the $\lambda - \epsilon$ plane for a fixed metallicity.
In Fig.~\ref{fig:chem_corner_plot}, we show this for a stellar metallicity of $[Z/H]_{\rm SSP} = -0.67$~dex and mask the region with $f_{\rm out} > 1$, together with iso-contours corresponding to $f_{\rm out} = 0.25, 0.5, 0.75, 0.85$, and $0.95$.
The tight $\lambda - \epsilon$ distribution lies between $f_{\rm out} = 0.85$ and $0.95$, implying that only $5$--$15\%$ of the produced metals are retained in the galaxy.
As for the degeneracy involving the infall metallicity $Z_{\rm inf}$, there is no strong one-to-one correlation with either $\lambda$ or $\epsilon$ separately, but together they show that a more metal-rich infall requires a higher $f_{\rm out}$.

The main outcome of the model is that two broad classes or ``families'' of solutions can reproduce the observed properties of the galaxy: a recent, short-timescale, metal-enriched gas infall event or an earlier, longer-timescale, metal-poor gas accretion episode.
The former scenario is more naturally associated with a gas-rich merger, whereas the latter is more consistent with accretion of gas from a cosmic filament.
A more detailed discussion of these possibilities is given in Sect.~\ref{subsec:origin}.

\section{Discussion}
\label{sec:discussion}

In this section, we investigate the most plausible formation scenario for the secondary disk in NGC~5717 in the context of its stellar populations, gas properties, and group environment.
We then  discuss the outcomes of the chemical evolution model for the star-formation process and feedback in the counter-rotating disk, and finally consider why such systems (brightest cluster or group galaxies, B(C/G)Gs, with embedded disks) are so rarely observed: whether we are catching a unique stage of B(C/G)G assembly or whether these systems are intrinsically rare. 

\subsection{Origin of the disk}
\label{subsec:origin}

For a deeper understanding of how cold gas accretion forms new kinematically decoupled components, large and statistically representative samples of such objects are required \citep{Bevacqua2022MNRAS.511..139B,Gasymov2025ApJS..281...19G,Bao2026arXiv260304044B}. 
However, NGC~5717 attracted our attention because of its unusually dense environment. 
A sparse environment is generally considered favorable for the long-term accretion and survival of externally acquired cold gas, since such structures are less likely to be disrupted by subsequent galaxy interactions \citep{Oosterloo2007A&A...465..787O}. 
This picture is also supported by galaxies with kinematically decoupled gaseous or stellar components: decoupled gas disks are frequently found in isolated S0 galaxies \citep{Katkov2014MNRAS.438.2798K} and polar-ring galaxies preferentially inhabit low-density environments \citep{Finkelman2012MNRAS.422.2386F,Savchenko2017AstL...43..146S}. 
In this context, NGC~5717 is particularly remarkable: it is the first known BGG in a non-fossil group to host such a structure and, moreover, one of the most massive galaxies known to host a counter-rotating stellar disk.
What, then, is the origin of the secondary component in this extreme case?
Three channels are possible: internal dynamical evolution of the pre-existing stellar disk, accretion of ex-situ stars, or retrograde accretion of external gas followed by in-situ star formation.
Below we show that only the latter scenario can explain the observed properties of the galaxy, with gas accretion from \textit{a cosmic filament} being the most plausible origin of the acquired material.

There are several ways to form a CR disk:
\begin{enumerate}
    \item Internal
    \begin{enumerate}
        \item bar dissolution / separatrix crossing and triaxial-to-axisymmetric evolution \citep{Evans1994ApJ...420L..67E,Collett1997MNRAS.285...49C}.
    \end{enumerate}
    \item External stellar accretion (ex situ stars)
        \begin{enumerate}
        \item dissipationless retrograde mergers \citep{DiMatteo2008A&A...477..437D};
        \item finely tuned major mergers that preserve or rebuild a disk \citep{Puerari2001Ap&SS.276..909P,Crocker2009MNRAS.393.1255C}.
    \end{enumerate}
    \item External retrograde gas acquisition, followed by in situ star formation \citep{Thakar1996ApJ...461...55T,Thakar1998ApJ...506...93T}
    \begin{enumerate}
        \item gas accretion from or a minor merger with a gas-rich dwarf companion \citep{Thakar1997ApJ...479..702T};
        \item tidal gas transfer during interactions with nearby galaxies \citep{Khim2021ApJS..254...27K,Silchenko2023Galax..11..119S};
        \item gas retained from a gas-rich major merger progenitor \citep{DiMatteo2008A&A...477..437D};        
        \item gas accretion along two intersecting cosmological filaments \citep{Algorry2014MNRAS.437.3596A}.
    \end{enumerate}
\end{enumerate}

\textbf{Internal scenario:} An internal formation scenario would predict similar stellar-population properties for the two components, whereas in NGC~5717 the CR disk is clearly younger and {\it more metal-poor}.

\textbf{Ex situ stars:} 
In the case of NGC~5717, this scenario cannot explain properties of its CR disk because the disk is young (Age$_{\rm CR} \sim 1.5$ Gyr) and already dynamically settled, whereas the required merger and relaxation timescale is of the order of 1--1.5~Gyr \citep{DiMatteo2008A&A...477..437D, Crocker2009MNRAS.393.1255C}.
This would imply that the merger event had to begin when the possible ``progenitor'' galaxy for the CR disk was only just being assembled, which appears unrealistic.
Another concern is the clear spatial and kinematic association of the ionized-gas emission with the dust rings, and the counter-rotating disk (see Figs.~\ref{fig:kin_maps}, \ref{fig:losvd_1d}, and \ref{fig:2comp_fit}). 
In a dry-merger scenario, which is by definition gas-poor, this connection is not straightforward to explain, since the gaseous and dusty structures would require either an additional source of cold gas or a later gas-accretion event that is not intrinsic to the dissipationless merger itself.
Finally, the stellar CR disk is dynamically cold ($V/\sigma \approx 3$), which is more naturally explained if it formed in situ from externally acquired gas after dissipative settling. 
In contrast, a secondary component assembled predominantly from ex situ stars in a dry merger would generally be expected to be dynamically hotter \citep{Read2008MNRAS.389.1041R, McCluskey2024MNRAS.527.6926M}.

\textbf{External gas and in-situ formed stars:} As mentioned above, the most likely scenario is in-situ star formation from externally accreted gas. 
Under the assumption that the present-day stellar and gaseous counter-rotating components are directly linked, the low metallicity of the secondary stellar disk ([Fe/H]$_{\rm CR} \approx -0.7$ dex) strongly disfavors a metal-rich progenitor: the mass--metallicity relation implies near-solar abundances for massive donors \citep{Curti2020MNRAS.491..944C,Mattolini2025A&A...703A...5M}, while negative radial gradients lower the abundance in their outskirts only to moderately subsolar values \citep{Sanchez2014A&A...563A..49S}.
Moreover, stars formed from such gas would be expected to become more metal-rich owing to further chemical enrichment during the evolution of the secondary component.
The total mass of the accreted gas must have been at least comparable to the present-day baryonic content of the secondary component, that is, $M_{\rm acc} \gtrsim M_{\rm CR}+M_{\rm gas} \sim 10^{10}\,M_\odot$, although its initial metallicity remains uncertain. 

The chemical evolution model (Sect.~\ref{subsec:chem_mod}) admits a large number of solutions, from which we identify two representative families with different physical interpretations: either a prolonged accretion of metal-poor gas ([Z/H]$_{\rm inf}\lesssim -1$ to $-1.5$), or a shorter accretion episode of mildly pre-enriched material ($-1.5 \lesssim$ [Z/H]$_{\rm inf} \lesssim -0.8$).
The latter could in principle be associated with an external galaxy, but a massive gas-rich progenitor would be expected to have substantially higher gas-phase metallicity, again following the galaxy mass-metallicity relation. 
A giant low-surface-brightness galaxy \citep{KuziodeNaray2004MNRAS.355..887K, Saburova2024ApJ...973..167S, Saburova2026ApJ...998...19S} or the combined accretion of several gas-rich dwarfs \citep{Egorova2021MNRAS.504.6179E, Egorova2026A&A...705A..45E} cannot be excluded a priori; however, both possibilities would require a rather specific sequence of accretion events involving these rare gas-rich and metal-poor progenitors.
In this context, long-timescale accretion from a cosmic filament appears to be the most plausible explanation, since cosmological gas accretion is expected to be metal-poor and can sustain extended in situ star formation over Gyr timescales \citep{Algorry2014MNRAS.437.3596A, SanchezAlmeida2014A&ARv..22...71S}. 
As an independent environmental constraint, we examined the cosmic-web morphology around NGC~5717.
Cosmic-web filaments trace ridges of the large-scale matter-density field, while nodes correspond to locations where several such ridges meet \citep{Kraljic2020MNRAS.491.4294K}.
Hydrodynamical simulations show that gas and satellites can be accreted anisotropically along these structures and that dense cold streams may penetrate hot halo atmospheres, particularly at high redshift \citep{Keres2005MNRAS.363....2K,Dekel2009Natur.457..451D,Algorry2014MNRAS.437.3596A,Taylor2018MNRAS.479..141T}.
At low redshift, \HI\ studies provide statistical evidence that galaxies near cosmic-web filaments may be replenished with cold gas, although this effect has primarily been established in low- and intermediate-density environments \citep{Kleiner2017MNRAS.466.4692K}. Although the cosmic-web geometry does not independently demonstrate the presence of a cold gaseous stream or determine its metallicity, the proximity of NGC~5717 to two intersecting filamentary branches provides independent geometrical support for a filament origin of the externally acquired gas that formed the CR disk.

In our preferred interpretation, filament accretion began about 2--3~Gyr ago and proceeded over a characteristic timescale of $\sim$0.3--1~Gyr, as inferred from the chemical evolution modeling.

\subsection{Star formation and feedback in the CR disk}
\label{subsec:SF_feautures}

The chemical evolution model described above is simplified in order to avoid over-parameterization given the limited number of available observational constraints.
Nevertheless, it reveals several important features of the star formation process in the disk.
As a first-order consideration, the surrounding elliptical component is large and massive, and the estimated escape velocity at the disk radius ($R\approx6$~kpc) is $V_{\rm esc} = \sqrt{2} V_{\rm CR} / \sin{i} \approx 400$~km~s$^{-1}$.
This value is comparable to the characteristic velocity of stellar-feedback-driven outflows, $V_{\rm out} \sim 300$--$500$~km~s$^{-1}$, although the SN-heated wind reaches velocities of order $10^3$~km~s$^{-1}$ \citep{Strickland2002ASPC..253..387S, Heckman2015ApJ...809..147H}. 
This motivated us to consider an ``accreting-box'' model of chemical enrichment, that is, a model with gas infall but without outflow (mass-loading factor $\lambda = 0$). 
The result is clear: such a model cannot reproduce the low metallicity of the CR disk independently of the infall parameters. 
The minimum allowed value is $\lambda \approx 0.11$, corresponding to an outflow driven solely by the SN-heated wind, which is consistent with the velocity estimates given above.

If outflows are indeed required by the model, the inferred metal-loss fraction $f_{\rm out}$ can be compared with available estimates (Sect.~\ref{subsec:chem_mod_app}).
In this context, the idea that the secondary component evolved as a nearly ``closed'' system inside a massive elliptical host can be directly tested.
In our framework, $f_{\rm out}$ is not constant but depends on the metallicity $Z(t)$ (Eq.~\ref{eq:f_out}); here we quote its value at the final metallicity, $Z = Z_{\rm CR} = Z_\odot 10^{{\rm [Fe/H]}_{\rm CR}}$.
For the adopted star formation efficiency $S = 2$~Gyr$^{-1}$, $f_{\rm out}$ lies between 86\% and 94\%, with only a weak dependence on $\lambda$ and/or $\epsilon$.
Even when varying $S$ (see below), the inferred metal-loss fraction changes only moderately and remains above 80\%.
Such values appear extremely high compared to those inferred for other galaxies.
For example, \citet{Christensen2018ApJ...867..142C} found in simulations that about 35--85\% of metals reside outside galaxy disks by $z=0$, with a strong anti-correlation with galaxy mass: the upper end of this range is reached only by low-mass dwarf systems, consistent with observational estimates from \citet{Kirby2011ApJ...742L..25K} and \citet{McQuinn2015ApJ...815L..17M}.
Although these quantities are not strictly identical to our effective $f_{\rm out}$, they suggest that values above 80\% are unusual for massive systems.
An extreme comparison case is provided by the starburst galaxy M82, where the X-ray-emitting hot wind has been argued to be consistent with carrying nearly all freshly produced metals from the ongoing starburst, with $V_{\rm out} \gtrsim 1000$~km~s$^{-1}$ \citep{Strickland2009ApJ...697.2030S, Peeples2011MNRAS.417.2962P}.
Therefore, if such large effective metal-loss fractions are indeed required by our model, the possible interpretation is that the secondary disk experienced a rapid starburst-like phase accompanied by a hot supernova-driven wind, which might have preferentially carried metal-rich ejecta \citep{Lochhaas2021MNRAS.504.3412L}.

The last important aspect of the chemical evolution model is the star formation efficiency, or, the reciprocal value, the gas depletion time in the galaxy.
Typical depletion times in star-forming galaxies are 1--2~Gyr \citep{Bigiel2011ApJ...730L..13B} or 0.5--2~Gyr \citep{Saintonge2011MNRAS.415...61S}.
However, in NGC~5717, models with long depletion times $t_{\rm dep} > 1 {\rm~Gyr} $, equivalent to $S < 1$~Gyr$^{-1}$, reproduce only a very limited set of solutions and only for high mass-loading factors, $\lambda > 0.8$. 
The reason is physical: in such an ``in''-efficient star-forming regime, building the stellar mass of the disk over a timescale of $\sim 1$--2 Gyr to yield $M_{\rm CR} \sim 10^{10}$\Msun, requires a large gas inflow, while at the same time the current gas mass is constrained to be $M_{g}(0) < 10^9$\Msun.
The latter is feasible only with some removal of excess gas, which within our model is possible only through outflows and therefore requires a large value of $\lambda$.
To reproduce solutions over the full considered range of $\lambda$ ($0 \le \lambda \le 1$), the star formation efficiency must be at least $S \approx 1.4$--$1.5$~Gyr$^{-1}$, or equivalently the depletion time must satisfy $t_{\rm dep} \lesssim 0.7$~Gyr, which already corresponds to a ``bursty'' regime.

This is why we adopted $S = 2$~Gyr$^{-1}$ as a representative value for NGC~5717, although more extreme starburst-like solutions with $S \gg 1$~Gyr$^{-1}$ formally exist and are limited mainly by the physics of star formation itself. 
More generally, values in the range of $1.4 \lesssim S \lesssim 5~{\rm Gyr}^{-1}$ reproduce the model well and recover the two possible families of inflow solutions discussed in Sect.~\ref{subsec:chem_mod_app}.
These values correspond to $0.2 \lesssim t_{\rm dep} \lesssim 0.7$~Gyr, which is firmly within the starburst regime \citep{Kamenetzky2017MNRAS.471.2917K,Kennicutt2021ApJ...908...61K}. 
In addition, the lack of strong gradients in the stellar population parameters (Age$_{\rm CR}$ and [Fe/H]$_{\rm CR}$; Fig.~\ref{fig:2comp_fit}) is consistent with the fast formation of the whole CR disk, without prolonged inside-out evolution \citep{Perez2013ApJ...764L...1P}. 

Taken together, these arguments suggest that the disk in NGC~5717 formed in a highly efficient, starburst-like regime, accompanied by a hot supernova-driven, metal-enriched outflow as observed in the starburst galaxy M82.

\subsection{NGC~5717: a rare system or a short-lived phenomenon?}

The next question is why we do not observe many other B(C/G)Gs with prominent secondary stellar kinematic components similar to that in NGC~5717. 
Is NGC~5717 intrinsically rare, or does it belong to a larger population of systems that are visible only during a relatively short and possibly unstable phase? 
This kind of object provides a test of scenarios involving external cold-gas supply and subsequent in situ star formation in massive central galaxies, independently of the final angular momentum of the newly formed component.

The sample of galaxies with CR disks presented by \citet{Gasymov2025ApJS..281...19G} contains only one BGG, NGC~5717, in intermediate- and large-group environments with more than 10 known members. 
A cross-match between the group catalog of \citet{Yang2007ApJ...671..153Y} and the MaNGA survey returned 249 B(C/G)Gs in groups with halo mass $M_h > 13.5$ and number of known members $N_{mem} > 10$, of which 231 are elliptical, lenticular, or transitional Sa galaxies (T-Type < 2; \citealt{Vazquez-Mata2025MNRAS.544.1056V}). 
Among them, only 45 galaxies show prominent emission lines, and only NGC~5717 (MaNGA-ID: 1-593328) shows a prominent secondary stellar component. 
Two other galaxies may host secondary co-rotating disks, although without strong spectroscopic confirmation (MaNGA-IDs: \href{https://manga.voxastro.org/visualiser/9038-12702}{1-199742} and \href{https://manga.voxastro.org/visualiser/10508-6101}{1-175865}). 
In addition, among the 48 \HI\ spectra available for the B(C/G)G subsample in MaNGA \citep{Masters2019MNRAS.488.3396M,Stark2021MNRAS.503.1345S}, cold gas was detected in only 10 galaxies; of these, two belong to the green valley, while the rest lie on the red sequence (using $\log$~sSFR~$< -10.8$ as the adopted threshold). 
Overall, we do not find evidence for a large population of B(C/G)Gs with prominent cold-gas reservoirs or ongoing/recent star formation, which suggests that systems like NGC~5717, hosting a prominent newly formed stellar component, are indeed rare.

In the case of NGC~5717, the luminosity-weighted age of the secondary component is Age$_{\rm CR} \sim$ 1.3--1.5 Gyr, while the chemical evolution model suggests that the galaxy could have hosted a substantial cold-gas reservoir for a relatively short period of $\sim$0.4--0.6 Gyr during the formation of the massive stellar component (Fig.~\ref{fig:chem_models}). 
This may represent an important signature of such systems. 
In real galaxies, however, this phase is likely to be more complex than in the idealized direct cold-gas accretion scenario assumed in the chemical evolution model. 
The externally accreted gas may require additional time to settle into a dynamically ordered configuration, and the relaxation of misaligned gaseous structures takes $t_{\rm relax} \sim 0.5$--$2$ Gyr \citep{vandeVoort2015MNRAS.451.3269V,Davis2016MNRAS.457..272D}. 
Hence, the timescale inferred from the chemical evolution model should be interpreted as an effective lower limit for the cold-gas-rich phase. 
In this sense, even the gaseous relaxation stage alone is expected to proceed on $\sim$Gyr timescales rather than representing a genuinely brief ``unique moment''.

We can attempt to assess the rarity of systems such as NGC~5717, since their formation likely requires the conjunction of several relatively independent conditions. 
A key factor is the probability of the required external cold-gas supply to a future BCG/BGG since such events are not ubiquitous in the assembly histories of massive galaxies. 
The recent work by \citet{Zhuang2026ApJ...999L..19Z} presented the first direct observation of a gas-rich minor merger onto a BCG at $z=0.25$, while estimates based on IllustrisTNG100 predict gas-rich accretion rates of the order of $\sim 0.6$~Gyr$^{-1}$, in good agreement with the toy model of \citet{Davis2019MNRAS.486.1404D}. 
However, these estimates apply only to galaxies with detected cold gas or evidence of ongoing accretion, which constitute only $\sim$10--20\% of the population, consistent with our estimate from the MaNGA survey discussed above. 
Taken together, these results indicate that such events are plausible, but far from guaranteed for every system. 

Moreover, a gas-accretion event alone does not guarantee the formation of a stable, long-lived secondary gaseous component. 
The accreted material must survive shocks, ram-pressure stripping, and angular-momentum loss, and then settle into a dynamically ordered gaseous configuration, a process that itself may proceed on $\sim$Gyr timescales \citep{vandeVoort2015MNRAS.451.3269V,Davis2016MNRAS.457..272D}. 
In addition, a dense group environment is generally not favorable for the long-term survival and dynamical settling of externally acquired cold gas, in contrast to the sparse environments where similar systems are more commonly found \citep{Oosterloo2007A&A...465..787O}. 
The cold gas must not only be accreted, but also survive long enough against heating or removal by AGN activity, the intra-group medium, and subsequent interactions. 

In this context, the evolutionary stage of the NGC~5717 group may be crucial for the formation of such a system.
Although the group is currently consistent with a virialized state, the epoch at which virial equilibrium was established remains unconstrained.
Since dynamical relaxation of a group-sized halo may require approximately one to two dynamical times, corresponding to a few Gyr, the group could have reached its present state only relatively recently.
This intermediate stage of group assembly may provide a favorable window for earlier external gas delivery, possibly from a cosmic filament (Sect.~\ref{subsec:origin}), followed by progressive virialization and dynamical stabilization of the group, with fewer strong disturbances capable of destroying the newly formed component. 
An indirect argument in favor of this picture is the absence of obvious tidal debris even in the deep CFHT $r$-band images, which may indicate the lack of a recent minor or major merger over the last $\sim$0.7--4 Gyr \citep{Mancillas2019A&A...632A.122M}. 
This timescale of a few Gyr is of the same order as that required for gas accretion, relaxation, and the formation of the secondary stellar component.

In summary, we propose that the rarity of NGC~5717-like systems is driven by a combination of several low-probability factors: external gas accretion onto a future B(C/G)G during a relatively short window in the group assembly history, followed by sufficiently calm evolution over several Gyr, allowing the gaseous component to relax and a secondary co-/counter-rotating stellar disk to form in situ.

\section{Conclusions}
We investigated the nature and origin of the counter-rotating stellar disk in NGC~5717 using MaNGA IFU spectroscopy, deep optical imaging, spectroscopy of the surrounding group from SDSS and DESI, and a simple chemical-evolution model. 
The main conclusions of this work are as follows:

\begin{itemize}[leftmargin=*, itemsep=0.4em, topsep=0.4em]
  \item NGC~5717 hosts a prominent dynamically cold counter-rotating stellar disk embedded in a dynamically hot elliptical component. 
  The galaxy is the brightest group galaxy in a system with a halo mass $\log M_h \approx 13.7$ and 27 previously known members, complemented by 16 additional candidate members identified from DESI spectroscopy within the group environment.
  The CR disk is spatially associated with the ionized gas component, has a young and metal-poor stellar population (Age$_{\rm CR} = 1.35~\pm~0.16$~Gyr, [Fe/H]$_{\rm CR} = -0.67~\pm~0.09$~dex), and is both extended and massive ($h_{\rm CR} = 2.43 \pm 0.02$~kpc, $\log M_{\star,{\rm CR}}/M_\odot = 9.9 \pm 0.1$).

  \item The observed properties of the secondary component indicate that the CR disk formed \textit{in situ} from externally accreted gas.
  This interpretation is consistent with the formation channel usually invoked for counter-rotating disks, although NGC~5717 is unusual in being located at the center of a group.
  Alternative scenarios, such as internal secular formation or the accretion of \textit{ex situ} stars, do not naturally explain the observed properties of the disk.
  
  \item Our preferred interpretation, based on the chemical-evolution model and the geometry of the surrounding cosmic web, is that the accreted material was supplied through prolonged accretion of metal-poor gas, plausibly associated with a cosmic filament, beginning about 2--3~Gyr ago.
  This scenario provides the most natural explanation for the low metallicity of the CR disk.

  \item The observed properties of the gas and the CR disk suggest that the secondary component formed in a highly efficient, burst-like regime of star formation.
  In our chemical-evolution framework, reproducing the stellar mass, metallicity, and low present-day gas content of the disk requires short depletion times together with substantial metal loss, consistent with a strong metal-enriched outflow driven by supernova feedback and sufficiently fast to escape the deep potential well of the massive elliptical host component.
  
  \item We suggest that the rarity of the population of large co-/counter-rotating stellar disks in B(C/G)Gs is due to the need for several conditions to coincide: external cold-gas accretion onto a future central galaxy, gas survival and settling during a relatively short window in the assembly of the surrounding group, and sufficiently undisturbed subsequent evolution for the newly formed disk to survive until the present day.
      
\end{itemize}

\begin{acknowledgements}
This research used the facilities of the Canadian Astronomy Data Centre operated by the National Research Council of Canada with the support of the Canadian Space Agency. The deep photometric imaging data are based on observations obtained with MegaPrime/MegaCam, a joint project of CFHT and CEA/DAPNIA, at the Canada-France-Hawai'i Telescope (CFHT), which is operated by the National Research Council (NRC) of Canada, the Institut National des Science de l'Univers of the Centre National de la Recherche Scientifique (CNRS) of France, and the University of Hawai'i. This research used data obtained with the Dark Energy Spectroscopic Instrument (DESI). DESI construction and operations are managed by the Lawrence Berkeley National Laboratory. This material is based upon work supported by the U.S. Department of Energy, Office of Science, Office of High-Energy Physics, under Contract No. DE–AC02–05CH11231, and by the National Energy Research Scientific Computing Center, a DOE Office of Science User Facility under the same contract. Additional support for DESI was provided by the U.S. National Science Foundation (NSF), Division of Astronomical Sciences under Contract No. AST-0950945 to the NSF’s National Optical-Infrared Astronomy Research Laboratory; the Science and Technology Facilities Council of the United Kingdom; the Gordon and Betty Moore Foundation; the Heising-Simons Foundation; the French Alternative Energies and Atomic Energy Commission (CEA); the National Council of Humanities, Science and Technology of Mexico (CONAHCYT); the Ministry of Science and Innovation of Spain (MICINN), and by the DESI Member Institutions: www.desi.lbl.gov/collaborating-institutions. The DESI collaboration is honored to be permitted to conduct scientific research on I’oligam Du’ag (Kitt Peak), a mountain with particular significance to the Tohono O’odham Nation. Any opinions, findings, and conclusions or recommendations expressed in this material are those of the author(s) and do not necessarily reflect the views of the U.S. National Science Foundation, the U.S. Department of Energy, or any of the listed funding agencies. 
Funding for the Sloan Digital Sky Survey IV has been provided by the Alfred P. Sloan Foundation, the U.S. Department of Energy Office of Science, and the Participating Institutions. SDSS-IV acknowledges support and resources from the Center for High-Performance Computing at the University of Utah. The SDSS website is www.sdss.org.
SDSS-IV is managed by the Astrophysical Research Consortium for the Participating Institutions of the SDSS Collaboration including the Brazilian Participation Group, the Carnegie Institution for Science, Carnegie Mellon University, the Chilean Participation Group, the French Participation Group, Harvard-Smithsonian Center for Astrophysics, Instituto de Astrof\'isica de Canarias, The Johns Hopkins University, Kavli Institute for the Physics and Mathematics of the Universe (IPMU) / University of Tokyo, the Korean Participation Group, Lawrence Berkeley National Laboratory, Leibniz Institut für Astrophysik Potsdam (AIP), Max-Planck-Institut für Astronomie (MPIA Heidelberg), Max-Planck-Institut für Astrophysik (MPA Garching), Max-Planck-Institut für Extraterrestrische Physik (MPE), National Astronomical Observatories of China, New Mexico State University, New York University, University of Notre Dame, Observat\'ario Nacional / MCTI, The Ohio State University, Pennsylvania State University, Shanghai Astronomical Observatory, United Kingdom Participation Group, Universidad Nacional Aut\'onoma de M\'exico, University of Arizona, University of Colorado Boulder, University of Oxford, University of Portsmouth, University of Utah, University of Virginia, University of Washington, University of Wisconsin, Vanderbilt University, and Yale University.
This work made use of Astropy,\footnote{\url{https://www.astropy.org}} a community-developed core Python package and an ecosystem of tools and resources for astronomy \citep{astropy:2013, astropy:2018, astropy:2022}.
The first author (DG) used ChatGPT (OpenAI) to assist with preparing and refining analysis and plotting scripts, and with language editing and stylistic revision of author-provided manuscript drafts. 
All outputs were reviewed and validated by the authors, who take full responsibility for the content of this work.
\end{acknowledgements}

%
  \bibliographystyle{aa} 
  \bibliography{bib} 
%
\appendix

\refstepcounter{section}
\label{app:morph_mismatch}

\twocolumn[{%
\centering
\includegraphics[width=0.98\textwidth]{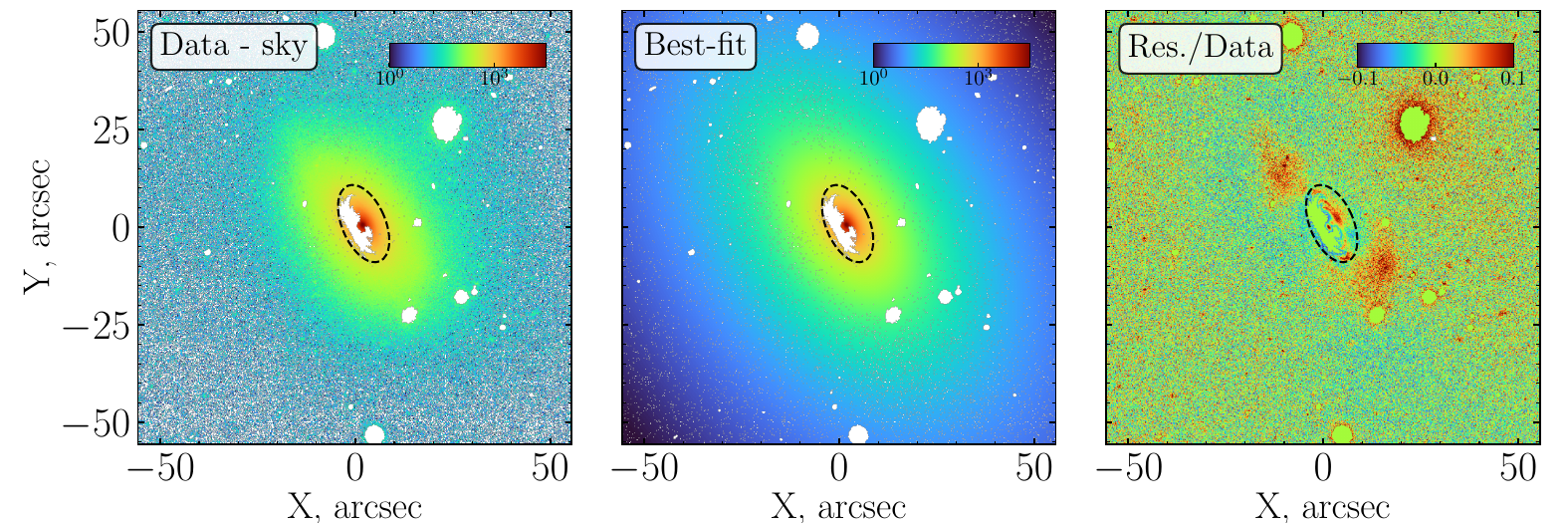}\par
\captionof{figure}{Two-dimensional photometric decomposition of NGC~5717 performed with \textsc{imfit}.
From left to right: the sky-subtracted CFHT $r$-band image, the best-fitting model, and the relative residual map, defined as $(\mathrm{Data}-\mathrm{Model})/\mathrm{Data}$.
The model consists of a point source, an exponential disk, and a S\'ersic component.
The black ellipse approximately marks the outer extent of the CR disk (see Fig.~\ref{fig:kin_maps}).
The data and model images are shown on a logarithmic scale, while the relative residuals are shown on a linear scale.
}%
\label{fig:imfit_test}\par
\vspace{1em}
}]

\section*{Appendix~\thesection: Morphological mismatch of NGC~5717}

Several catalogs classify NGC~5717 as a lenticular or early-type spiral galaxy. \citet{Kochanek2001ApJ...560..566K} reported a numerical morphological type of $T = -0.4 \pm 1.5$, placing the galaxy near the transition between lenticular and early-type Sa galaxies.
The HyperLEDA database \citep{Paturel2003A&A...412...45P} gives $T = 0.0 \pm 2.0$.
The MaNGA Visual Morphology catalog \citep{Vazquez-Mata2025MNRAS.544.1056V} classifies NGC~5717 as an Sa galaxy.
In contrast, throughout this work we interpret NGC~5717 as an elliptical galaxy hosting an embedded counter-rotating disk and refer to its dominant old stellar component as the elliptical component.
This apparent morphological discrepancy could affect the interpretation of our results and potentially confuse the reader.
We therefore perform two simple, independent tests to assess the nature of the dominant stellar component.

As a first, kinematic test, we consider the relative importance of ordered and random stellar motions.
From the recovered stellar line-of-sight velocity distribution, we measure a projected rotation velocity of $107 \pm 3$ \kms\ for the main component and an approximately constant velocity dispersion of $181 \pm 3$ \kms\ at large radii.
These values yield $V / \sigma \approx 0.6$, indicating that random stellar motions dominate over ordered rotation.
The main component is therefore dynamically hot, supporting its interpretation as a spheroidal rather than a dynamically cold disk (lenticular or early-type spiral) component.

As a second, photometric test, we estimate the S\'ersic index of the extended stellar component.
We fit the CFHT $r$-band image with \textsc{imfit} \citep{Erwin2015ApJ...799..226E} using three components: a point source to account for the AGN, an exponential profile to describe the embedded counter-rotating disk, and a S\'ersic profile for the extended stellar component.
Sources surrounding the galaxy were masked using \textsc{SExtractor} \citep{Bertin1996A&AS..117..393B}.
To reduce the effect of dust lanes, we adopted a two-step fitting procedure: at first, we fit the entire galaxy with the described three-component model, and in the second step, we additionally masked pixels with relative residuals $(\mathrm{Data}-\mathrm{Model})/\mathrm{Data} < -0.1$, corresponding to flux deficits exceeding 10\% relative to the initial model, and repeated the fit.
The final decomposition is shown in Fig.~\ref{fig:imfit_test}.
The extended component has a S\'ersic index of $n = 4.02$, close to the $n = 4$ de Vaucouleurs profile characteristic of classical ellipticals.
Together with its low $V / \sigma$, this result supports our interpretation of the dominant old stellar component as a dynamically hot spheroid.

\end{document}